\documentclass[twocolumn,amsmath,amssymb,prx]{revtex4-2}
\usepackage{latexsym}
\usepackage{graphicx}
\usepackage{psfig}
\usepackage{color}
\usepackage{verbatim}
\usepackage{multirow}
\usepackage[letterpaper, margin=1in]{geometry}
\usepackage{siunitx}
\usepackage[capitalise]{cleveref}
\usepackage{circledsteps}
\usepackage{lipsum}
\usepackage{physics}
\usepackage{mathtools}

\begin{document}

\preprint{}
\title{Microwave Andreev bound state spectroscopy in a semiconductor-based Planar Josephson junction}
\author{Bassel Heiba Elfeky$^{1}$}
\author{Krishna Dindial$^{1}$}
\author{David S. Brandão$^{2}$}
\author{Barış Pekerten$^{2}$}
\author{Jaewoo Lee$^{1}$}
\author{William M. Strickland$^{1}$}
\author{Patrick J. Strohbeen$^{1}$}
\author{Alisa Danilenko$^{1}$}
\author{Lukas Baker$^{1}$}
\author{Melissa Mikalsen$^{1}$}
\author{William Schiela$^{1}$}
\author{Zixuan Liang$^{1}$}
\author{Jacob Issokson$^{1}$}
\author{Ido Levy$^{1}$}
\author{Igor \v{Z}uti\'{c}$^{2}$}
\author{Javad Shabani$^{1}$} \email[]{jshabani@nyu.edu}

\affiliation{$^{1}$Center for Quantum Information Physics, Department of Physics, New York University, New York 10003, USA}
\affiliation{$^{2}$Department of Physics, University at Buffalo, State University of New York, Buffalo, New York 14260, USA}

\date{\today}

\begin{abstract}
By coupling a semiconductor-based planar Josephson junction to a superconducting resonator, we investigate the Andreev bound states in the junction using dispersive readout techniques.
Using electrostatic gating to create a narrow constriction in the junction, our measurements unveil a strong coupling interaction between the resonator and the Andreev bound states. This enables the mapping of isolated tunable Andreev bound states, with an observed transparency of up to 99.94\% along with an average induced superconducting gap of $\sim \SI{150}{\micro eV}$. Exploring the gate parameter space further elucidates a non-monotonic evolution of multiple Andreev bound states with varying gate voltage. Complimentary tight-binding calculations of an Al-InAs planar Josephson junction with strong Rashba spin-orbit coupling provide insight into possible mechanisms responsible for such behavior. Our findings highlight the subtleties of the Andreev spectrum of Josephson junctions fabricated on superconductor-semiconductor heterostructures and offering potential applications in probing topological states in these hybrid platforms. 

\end{abstract}

\maketitle

\section{Introduction}

Advancements in material growth have made it possible to create superconductor-semiconductor heterostructures with a strong proximity effect, leading to the development of novel, voltage-tunable, wafer-scale superconducting circuit elements \cite{deLange2015, Larsen_PRL, Luthi2018, kringhoj2018, Casparis2018, CasparisBenchmarking, yuan2021, kringhoj_parity2020, danilenko2022, hertel2022, Maxim17, Casparis2018, casparis2019, Scarlino2019, Borjans2020, sardashti2020, Burkard2020, phan2022, strickland_superconducting_2023, strickland_characterizing_2024}. Unlike conventional tunnel Josephson junctions, the supercurrent in a superconductor-semiconductor hybrid junction is carried by Andreev bound states (ABSs). This provides new degrees of freedom which can be explored in these hybrid devices. Beyond conventional tunable superconducting qubits, these material systems have been used to develop hybrid qubits where information is encoded in the spin of a quasiparticle in an ABS, the so-called superconducting-spin qubit \cite{Hays2020,hays2021,pita-vidal_direct_2023}. Josephson junctions fabricated on superconductor-semiconductor heterostructures, such as Al-InAs structures, have also been studied for their potential application in topological fault-tolerant quantum computing \cite{mayer2019anom, FornieriNature2019, dartiailh_phase2021, banerjee_signatures_2023}. Probing the rich physics exhibited by the ABSs in these Josephson junctions (JJs) is critical in characterizing their topological properties, utilize them as Andreev spin qubits and understand the loss mechanisms in different superconducting circuits elements fabricated on these hybrid heterostructures. The majority of probing techniques have been focused on tunneling spectroscopy to detect localized ABSs \cite{fornieri_evidence_2019, ren_topological_2019, poschl_nonlocal_2022, banerjee_signatures_2023} in Al-InAs nanowires and planar JJs. A promising parity-preserving alternative to tunneling spectroscopy is microwave spectroscopy using circuit quantum electrodynamics (cQED) techniques which provide higher energy resolution and access to fast time-sensitive dynamics. cQED techniques have been successfully used to probe Andreev bound states in nanowires \cite{van_woerkom_microwave_2017, Tosi2019, wesdorp2022, matute-canadas_signatures_2022, fatemi_microwave_2022, zellekens_microwave_2022} and two-dimensional electron gas (2DEG) planar junctions \cite{chidambaram2022, hinderling_flip-chip-based_2023, hinderling_direct_2024}. 

In this work, we perform microwave spectroscopy of ABSs by electrostatically defining a narrow constriction in a wide planar Al-InAs JJ. By coupling the ABSs to a superconducting resonator, we demonstrate strong coupling mediated by virtual photon exchange between the resonator and the ABSs. Using two-tone spectroscopy measurements, we probe the flux and gate tunability of the Andreev spectrum and resolve individual and multiple ABS pair transitions with ABSs that exhibit near-unity transparency. The experimental trends are further discussed within a theoretical framework of tight-binding calculations.  

\section{Device design and concept}

The JJ devices considered here are fabricated on epitaxial Al-InAs heterostructure grown via molecular beam epitaxy \cite{Shabani2016, Kaushini2018, Yuan2020, strickland2022} on a \SI{500}{\micro m} thick InP substrate. The weak link of the JJ is an InAs 2DEG formed in a near-surface quantum well and contacted \textit{in-situ} with a thin Al film. A phosphoric-based III-V wet etch is used to define the JJ area and superconducting loop. After the junction gap is etched with an Al wet etch, $\sim \SI{100}{\nano m}$ of Nb is sputtered to form the microwave circuit. This is followed by the deposition of a blanket layer of $\sim \SI{40}{\nano m}$ AlO$_{x}$ using atomic layer deposition which acts as a gate dielectric. Finally, a patterned layer of Nb gates is sputtered. Further details on the growth and fabrication are provided in Appendix A.

\begin{figure*}[ht!]
    \centering
    \includegraphics[width=1.0\textwidth]{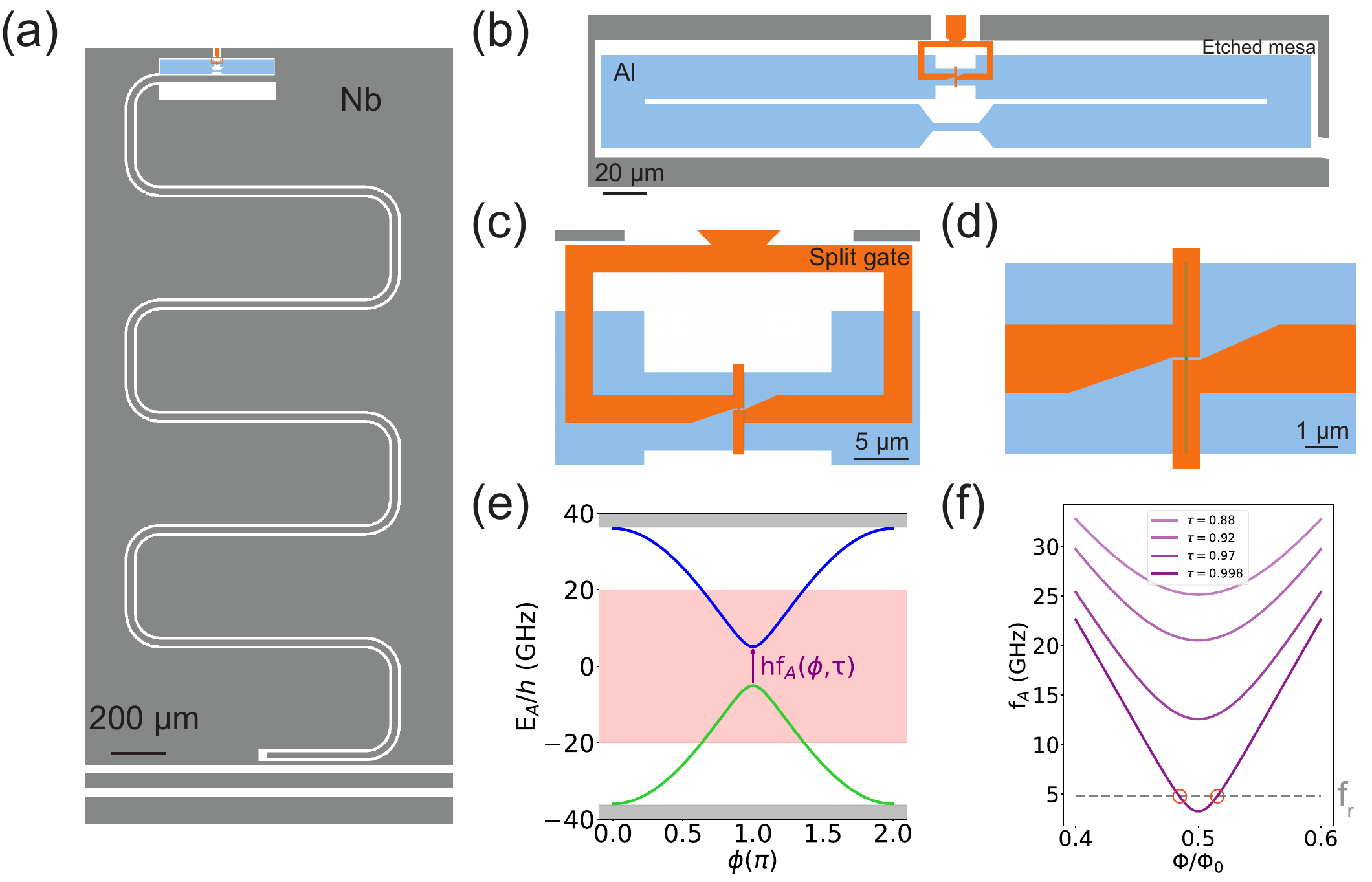}
     \caption{\textbf{(a)} Schematic of the device design where a superconducting loop with an Al-InAs junction is inductively coupled to a coplanar waveguide resonator. \textbf{(b)} A gated Al-InAs planar Josephson junction embedded in a superconducting loop. We confine the width of the loop in the bottom side of the superconducting loop to limit the amount of supercurrent flowing in the loop. The superconducting loop length, width and loop-resonator separation are designed to be $L_{l} = \SI{300}{\micro m}$, $ W_{l} = \SI{40}{\micro m}$ and $d = \SI{5}{\micro m}$, respectively. \textbf{(c)}, \textbf{(d)} Split gate designed to electrostatically define and tune a narrow constriction in the junction. The microwave circuit made of Nb is presented in grey, Al forming the superconducting loop and the junction electrodes in light blue,
     junction gap in green and the Nb used for the split gate in orange. \textbf{(e)} Energy of the positive (blue) and negative (green) branches of an ABS corresponding to a mode with $\Delta=\SI{150}{\micro eV}$ and $\tau = 0.98$. A transition between the negative and positive branches, represented with a purple arrow, requires an energy $hf_{A}(\phi,\tau)$. The energy space highlighted in red represents the accessible drive frequency range limited by the range of our microwave generator. \textbf{(f)} The excitation spectrum represented as the transition frequency $f_{A}$ corresponding to a transition between the negative and positive branches of an Andreev bound state with different transparencies and $\Delta=\SI{150}{\micro eV}$ as a function of applied flux $\Phi/\Phi_{0}$ where $\Phi/\Phi_{0} = \phi/2$. The grey dashed line correspond to the resonant frequency of the resonator $f_{r}$. The intersection of $f_{A}$ for a near-unity transparency mode with $f_{r}$ is represented by orange circles.
     }
    \label{fig:uwavespec_design}
\end{figure*}

\cref{fig:uwavespec_design}(a)-(d) presents schematics of the device design. The device consists of an Al-InAs Josephson junction embedded in a superconducting loop that is inductively coupled to a coplanar waveguide superconducting $\lambda/4$ resonator. Both the junction and the superconducting loop are fabricated on the Al-InAs heterostructure, while the Nb microwave circuit resides on the etched mesa. The resonator is then capacitively coupled to a common transmission line characterized by an external quality factor $Q_\mathrm{ext} \approx 9800$. The junction is defined to be $L \approx \SI{5}{\micro m}$ long (along the superconducting electrodes) and the normal region is $W_{N} \approx \SI{100}{\nano m}$ wide. The junction is equipped with a gate, referred to as a ``split gate'', that covers the junction area except for a small $L_{con}\sim \SI{100}{\nano m}$ long region in the middle of the junction, as illustrated in \cref{fig:uwavespec_design}(c)-(d), allowing to electrostatically define a tunable narrow constriction. By applying a magnetic flux $\Phi$ through the superconducting loop, the phase difference across the JJ is varied. On the bottom side of the superconducting loop, the width of the loop is constricted (\cref{fig:uwavespec_design}(b)) to limit the maximum supercurrent flowing in the loop. This allows for a stable phase drop across the junction $\phi \approx 2\pi \Phi/\Phi_{0}$ where $\Phi_{0} = h/2e$ is the superconducting flux quantum. Using $M = \frac{\mu_{0}}{2\pi}L_{l} \ln{(d+\frac{W_{l}}{d})}$, the mutual inductance ($M$) between the resonator and the superconducting loop can be estimated to be $\sim\SI{140}{pH}$ where $L_{l}$, $W_{l}$, $d$ is the superconducting loop length, width and loop-resonator separation, respectively. The resonator response, which reflects the ABS dynamics in the junction, is measured at a temperature of $T = \SI{15}{\milli K}$ through the transmission response (\textit{S$_{21}$}) of a readout probe tone applied through the transmission line. The measurement setup is expanded on in Appendix B.

\begin{figure*}[ht!]
    \centering
    \includegraphics[width=1.0\textwidth]{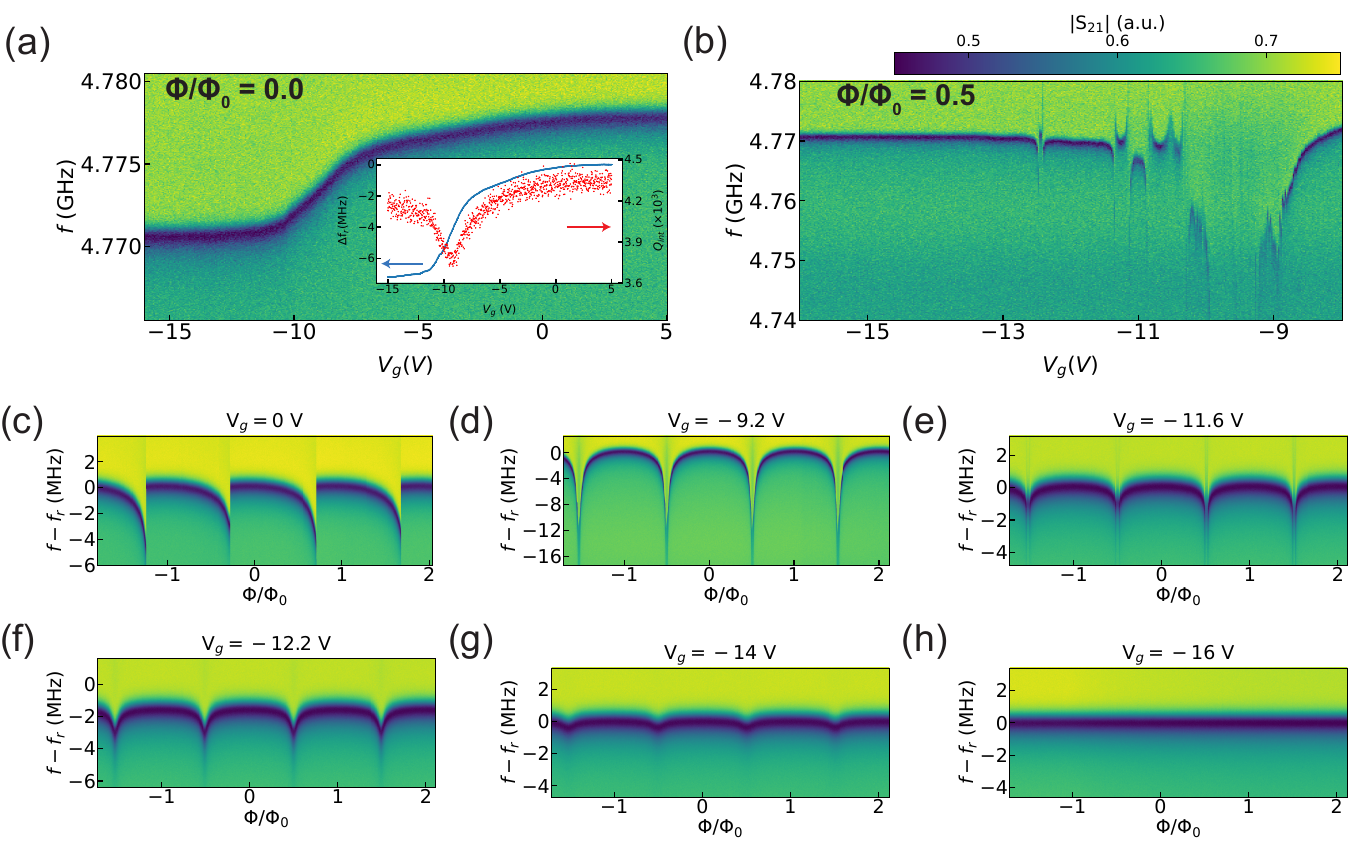}
     \caption{Resonator response to gate-tunability of the junction shown as the magnitude of the transmission coefficient $|S_{21}|$ as a function of the readout frequency $f$ and applied gate voltage $V_{g}$ at \textbf{(a)} $\Phi/\Phi_{0} = 0.0$ and \textbf{(b)} $\Phi/\Phi_{0} = 0.5$. Inset: Extracted change in resonant frequency $\Delta f_{r} = f_{r}(V_{g}) - f_{r}(V_{g} = \SI{0}{V})$ (blue) and internal quality factor $Q_{int}$ (red) as a function of $V_{g}$. \textbf{(c)-(h)} Flux modulation of the resonant frequency at different $V_{g}$.
     }
    \label{fig:uwavespec_tun}
\end{figure*}

Supercurrent in semiconductor-based JJs is carried by electrons and holes in conduction channels mediated by Andreev reflection \cite{beenaker1992}. Coherent processes of Andreev reflection result in the formation of sub-gap fermionic ABSs, where each channel forms a pair of ABSs. In its simplest form, the energy of the ABSs is expressed as:

\begin{equation}
    E_{A}^{\pm}(\phi) = \pm\Delta \sqrt{1-\tau\sin^2(\phi/2)}
    \label{eq:abs_pm}
\end{equation}

where $\phi$, $\tau$, and $\Delta$ are the phase difference across the junction, transparency, and superconducting gap, respectively. Typically, at low temperatures, the negative branch of the ABSs with energy $E_A^-$ are occupied, and the positive branch with energy $E_A^+$ are unoccupied. Driving a transition between the negative and positive branches of the ABS, as illustrated in \cref{fig:uwavespec_design}(e) for $\Delta=\SI{150}{\micro eV}$, then requires an energy $hf_{A}(\phi,\tau) = 2 |E_{A}|$. \cref{fig:uwavespec_design}(f) plots $f_{A} (\phi)$ corresponding to a pair transition between the negative and positive Andreev bound states for different $\tau$. 
We note that in reality the shape and depth of an ABS can have contributions from various effects rather than just the transparency. Since the positive and negative branches of an ABS carry current in opposite directions, a transition between the positive and negative bound states results in a change in the total supercurrent carried by the junction. The resulting change in supercurrent, and the corresponding change in the junction inductance, can be readout using standard dispersive measurement techniques \cite{van_woerkom_microwave_2017, Tosi2019, wesdorp2022, matute-canadas_signatures_2022, fatemi_microwave_2022, zellekens_microwave_2022, chidambaram2022, hinderling_flip-chip-based_2023, hinderling_direct_2024}. Since the resonator and superconducting loop are coupled inductively, the resonator-ABS coupling is mediated through phase fluctuations as theoretically outlined in Ref.\cite{park_adiabatic_2020}.

\section{Device characterization}

We present the resonator response in the form of the magnitude of the transmission coefficient $|S_{21}|$ measured with a vector network analyzer with a readout power corresponding to $\approx 2-5$ photons; the readout power dependence of the resonator response is discussed in Appendix C. \cref{fig:uwavespec_tun}(a) presents the resonator response as a function of $V_{g}$ at $\Phi/\Phi_{0} = 0$. As $V_{g}$ decreases, $f_{r}$ is seen to decrease slightly starting at $V_{g} = \SI{0}{V}$. This is due to a decrease in carrier density in the junction area covered by the split gate resulting in a decrease in supercurrent in that region. Starting $V_{g} = \SI{-7}{V}$, the number of modes under the split gate start to become heavily suppressed, resulting in a significant decrease in supercurrent, corresponding to an increase in the Josephson inductance $L_{J}$, reflected in the sharp decrease in $f_{r}$. Roughly around $V_{g} = \SI{-11}{V}$, the modes under the gates are completely suppressed and the narrow constriction between the split gates is defined. Decreasing the gate voltage further then electrostatically tunes the supercurrent in the constriction via the lateral dispersion of the electric field from the split gates. Beyond $V_{g} \approx \SI{-14.3}{V}$, the supercurrent in the constriction vanishes. The trend of $f_{r}$ with $V_{g}$, shown in the inset of \cref{fig:uwavespec_tun}(a), is consistent with the gradual suppression of lateral modes in the junction and the definition of a narrow constriction modes that are tunable with the split gates. The dependence of the internal quality factor $Q_{int}$ on $V_{g}$ shows a decrease in $Q_{int}$ in the $V_{g}$ parameter space corresponding to the constriction definition; a potential explanation for such dependence of $Q_{int}$ is discussed further in Appendix D.

The periodic modulation of the resonant frequency $f_{r}$ as a function of a magnetic flux $\Phi$ threading the loop is presented in \cref{fig:uwavespec_tun}(c)-(h), where the field is swept from negative to positive, for different values of $V_{g}$. 
For $V_{g} = \SI{0}{V}$, shown in \cref{fig:uwavespec_tun}(c), the flux oscillations are seen to exhibit hysteretic modulation with jumps around $\Phi/\Phi_{0} = 0.5$. This can be attributed to the presence of a finite loop inductance $L_{loop}$ with respect to the junction inductance $L_{J}$ resulting in phase-slips \cite{masluk_microwave_2012}. At negative values of $V_{g}$, where the supercurrent is suppressed and $L_{J}$ is higher, $L_{J}$ becomes large enough with respect with $L_{loop}$ resulting in flux modulation that is not hysteretic and the modulation becoming continuous as seen in \cref{fig:uwavespec_tun}(d)-(g). The cases shown in \cref{fig:uwavespec_tun}(e)-(g) correspond to a gate parameter space where supercurrent underneath the gates is completely suppressed and the supercurrent is mainly confined to the constriction. In that parameter space, the gate tunes the modes in the constrictions which results in the flux modulation varying with $V_{g}$. When the gate voltage  is decreased further the modulation is seen to become less drastic (\cref{fig:uwavespec_tun}(g)) and eventually reaches a point where the supercurrent is completely suppressed in the constriction, resulting in the absence of flux modulation as shown in \cref{fig:uwavespec_tun}(h). The gate dependence of the resonator at $\Phi/\Phi_{0} = 0.5$ is plotted in \cref{fig:uwavespec_tun}(b). Several patterns of vacuum Rabi splitting are observed as the resonator strongly interacts with one or more highly transparent ABS, where $f_{A}$ of the ABS is $\approx 2E_{A}/h$, as they are tuned by $V_{g}$. The resonance's visibility is compromised between $V_{g} = \SI{-9.5}{V}$ and $V_{g} = \SI{-12}{V}$, possibly due to a large number of possible transitions between ABSs.

\section{ABS-resonator coupling}

\begin{figure*}[ht!]
    \centering
    \includegraphics[width=1.0\textwidth]{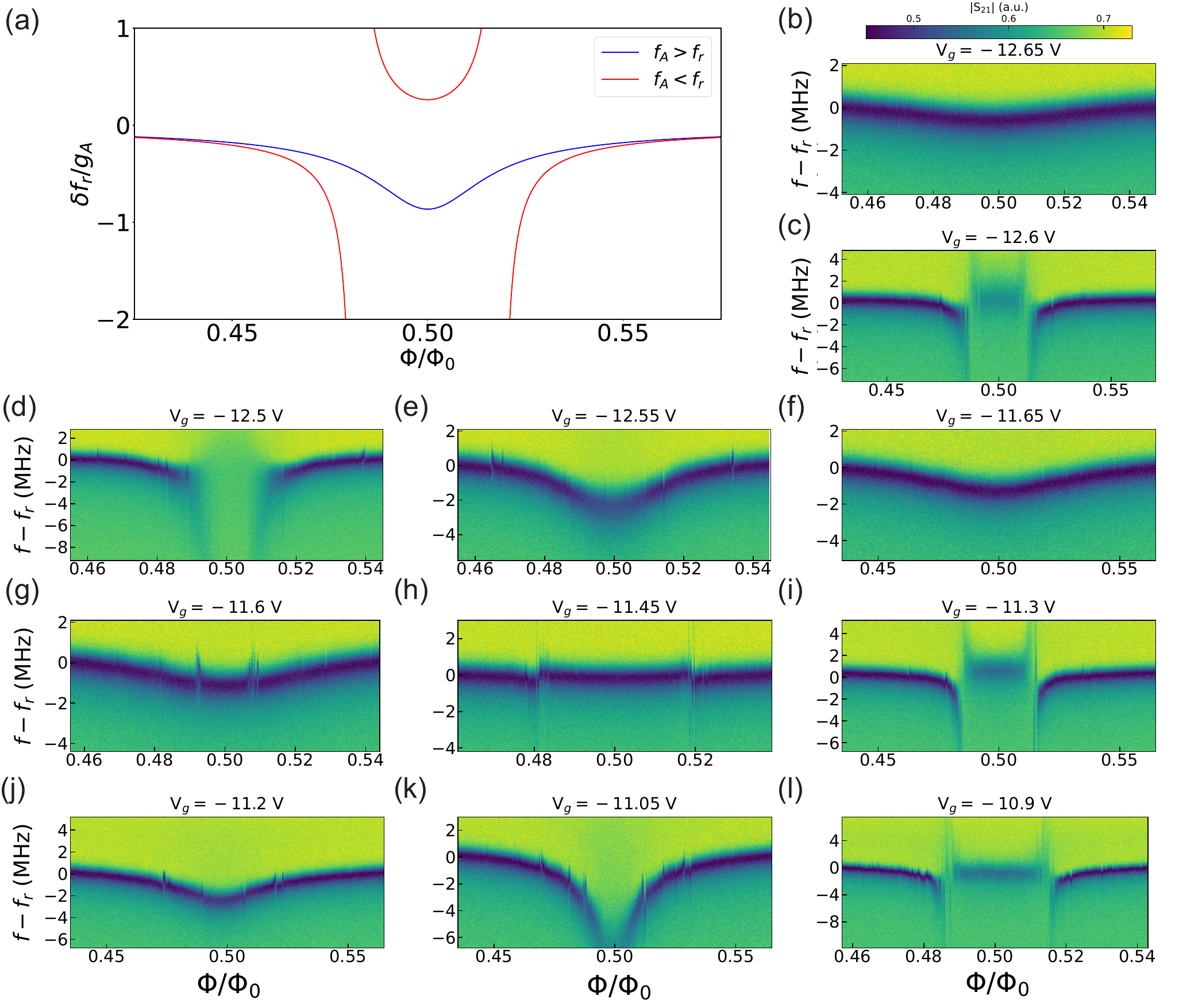}
     \caption{\textbf{(a)} Shift in resonant frequency normalized by the coupling rate $g_{A}$ 
     as a function of $\Phi/\Phi_{0}$ for two cases: 1) a mode with transparency $\tau = 0.993$ corresponding to $f_{A}>f_{r}$ for all values of $\Phi/\Phi_{0}$ and 2) a mode with $\tau = 0.999$ corresponding $f_{A}<f_{r}$ for a range of values of $\Phi/\Phi_{0}$. The superconducting induced gap is taken to be $\Delta = \SI{150}{\micro eV}$ and the resonant frequency to be $f_{r} = \SI{4.777}{\giga Hz}$.
     \textbf{(b)-(l)} Transmission coefficient $|S_{21}|$ as a function of the readout frequency $f$ and applied flux $\Phi$ around $\Phi/\Phi_{0} = 0.5$ for different $V_{g}$ values. 
     }
    \label{fig:uwavespec_ac}
\end{figure*}


In the presence of a single transparent mode in the constriction, the ABS-resonator coupling can be described by a resonator that is coupled to a two-level system consisting of a single pair of Andreev bound states (states $\Psi_{+}$ and $\Psi_{-}$ with energy $E_{A}^{+}$ and $E_{A}^{-}$) where  the excitation energy is then $f_{A}$. The resonator's response depends on the dispersion of $f_{A}(\tau,\phi)$ relative to $f_{r}$. As outlined in Ref.\cite{park_adiabatic_2020}, the ABS-resonator coupling can be described in the dispersive regime and the adiabatic regime. In the adiabatic regime, where $f_{A}$ and $f_{r}$ are strongly detuned, the resonator's response corresponds to the renormalization of the effective resonator inductance by that of the ABSs where the ABS inductance follows $L_{A} = (\Phi_{0}/2\pi)^2 (d^{2}E_{A}/d\phi^{2})^{-1}$. On the other hand, in the dispersive regime, the detuning between the resonator and ABSs is small enough to support coupling mediated through virtual photon exchange while still being larger than the coupling strength. The response of the resonator in the dispersive regime can be described by the Jaynes-Cummings (JC) model \cite{koch_charge-insensitive_2007} typically used to describe resonator-qubit coupling. Using the JC model, the shift in the resonant frequency can be described as $\chi_{A} \approx  \mp \frac{g^2_{A}}{2\pi (f_{A}-f_{r})}$ \cite{park_adiabatic_2020} where $g_{A}$ is the coupling strength between $\Psi_{-}$ and $\Psi_{+}$ which includes contributions from the coupling to the resonator. In \cref{fig:uwavespec_ac}(a), we present an example of the resonator response when $f_{A}>f_{r}$ (blue line) where the shift in resonant frequency follows the resonator's inductance being renormalizied by that of the ABS seen as a minimum in $f_{r}$ at $\Phi/\Phi_{0} = 0.5$. For a high transparency ABS, where $f_{A}<f_{r}$ for some range of $\Phi/\Phi_{0}$, the resonator response (red line) exhibits characteristic anticrossing, typical of the JC model. The avoided crossings occur at values of $\Phi/\Phi_{0}$ where $f_{A}\approx f_{r}$, as outlined in orange circles in \cref{fig:uwavespec_design}(f), causing $\chi_{A}$ to diverge and is an indication of a virtual resonator-ABS photon exchange that induces a push in the resonant frequency of the resonator. The shape of the resonator response depends mainly on $g_{A}$, $\Delta$ and $\tau$ of the ABS.

In \cref{fig:uwavespec_ac}(b)-(l), we present the flux dispersion of the resonance zoomed in around half-flux at different $V_{g}$ values where the flux dispersion of $f_{r}$ and its shape is seen to vary drastically with $V_{g}$. This implies the presence of ABSs in the constriction area that are gate-tunable. For the cases where the ABS transparency is not high enough for $f_{A} (\tau,\phi)$ to cross $f_{r}$, i.e. $V_{g} = -12.65, -12.55, -11.65, -11.2,\SI{-11.05}{V}$, $f_{r}$ continuously changes exhibiting a minimum at $\Phi/\Phi_{0} = 0.5$ resembling the $f_{A}>f_{r}$ case shown in \cref{fig:uwavespec_ac}(a). For the $V_{g}$ cases where an ABS is transparent enough for $f_{A}(\tau,\phi)$ to cross $f_{r}$, i.e. $V_{g} = -12.6, -12.5, -11.45, -11.3,\SI{-10.9}{V}$, prominent avoided crossings corresponding are observed similar to the $f_{A}<f_{r}$ case shown in \cref{fig:uwavespec_ac}(a). 
For $f_{A}$ to reach $f_{r} = \SI{4.777}{\giga Hz}$, an ABS transparency $\tau\gtrsim0.9957$ is required based on the simple model provided by \cref{eq:abs_pm}, assuming an induced superconducting gap $\Delta = \SI{150}{\micro eV}$. The value of $\Phi/\Phi_{0}$ at which the avoided crossing occurs at can be used to infer the ABS transparency to be $\tau \approx$ 0.9977, 0.99795, 0.99747 (assuming $\Delta = \SI{150}{\micro eV}$) for $V_{g} = \SI{-12.6}{V}$, \SI{-11.3}{V} and \SI{-10.9}{V}, respectively. In addition to the prominent avoided crossings, the flux dispersion displayed in \cref{fig:uwavespec_ac}(b)-(l) exhibits faint peaks/dips that likely correspond to multi-photon processes as discussed in Appendix E.


\begin{figure*}[ht!]
    \centering
    \includegraphics[width=1.0\textwidth]{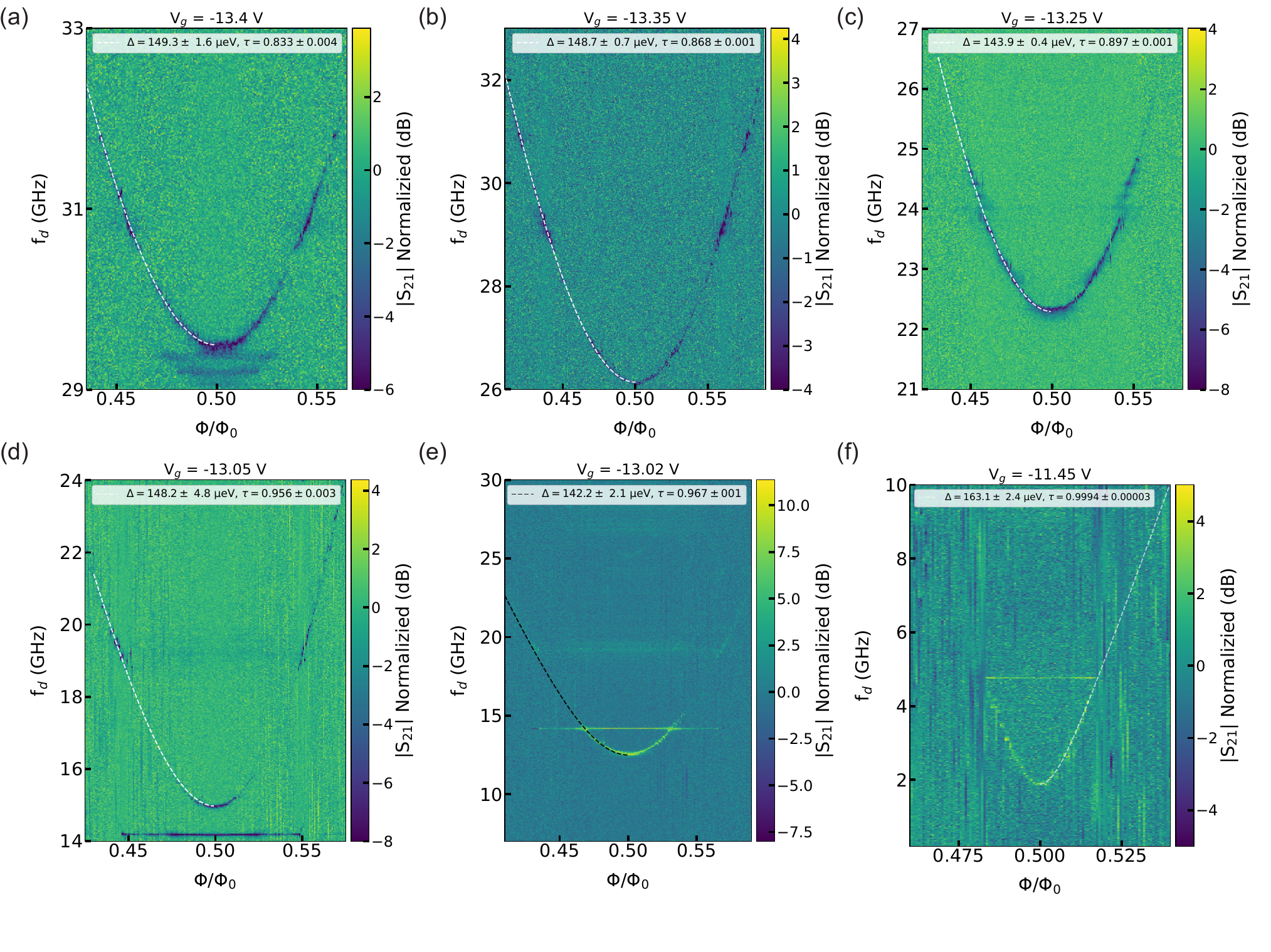}
     \caption{Transmission coefficient $|S_{21}|$ at a set readout frequency slightly offset $f_{r}$ as a function of drive tone frequency $f_{d}$ and applied flux $\Phi/\Phi_{0}$ for specific $V_{g}$ values where a single isolated ABS is observed. Fits of the parabola to $f_{A} = 2|E_{A}|/h$, where $E_{A}$ follows \cref{eq:abs_pm}, are shown as dashed lines along with the extracted values for the superconducting gap $\Delta$ and the transparency $\tau$.
     }
    \label{fig:uwavespec_spec_sm}
\end{figure*}

\section{Andreev spectroscopy}

\begin{figure*}[ht!]
    \centering
    \includegraphics[width=1.0\textwidth]{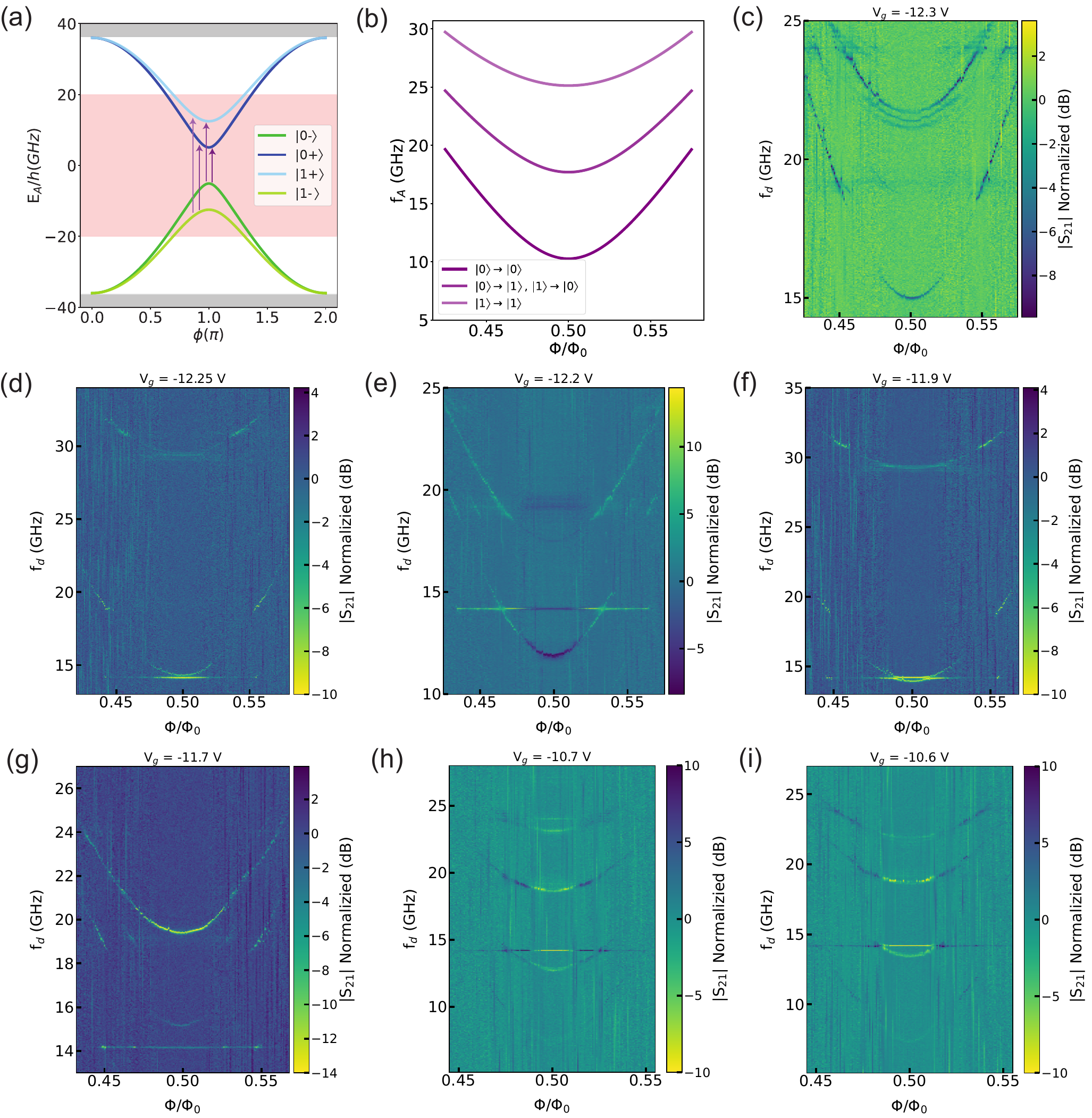}
     \caption{Positive (blue) and negative (green) energy branches of two Andreev bound states with $\Delta=\SI{150}{\micro eV}$ and $\tau = 0.98$ (dark) and $\tau = 0.88$ (light). The allowed transitions between the modes are represented by purple arrows. \textbf{(b)} Excitation spectra of $f_{A}$ corresponding to each transition outlined in (a). \textbf{(c)}-\textbf{(i)} $|S_{21}|$ as a function of drive tone frequency $f_{d}$ and flux $\Phi/\Phi_{0}$ for specific $V_{g}$ values where several transitions are observed.
     }
    \label{fig:uwavespec_spec_mm}
\end{figure*}

\begin{figure*}[ht!]
    \centering
    \includegraphics[width=1.0\textwidth]{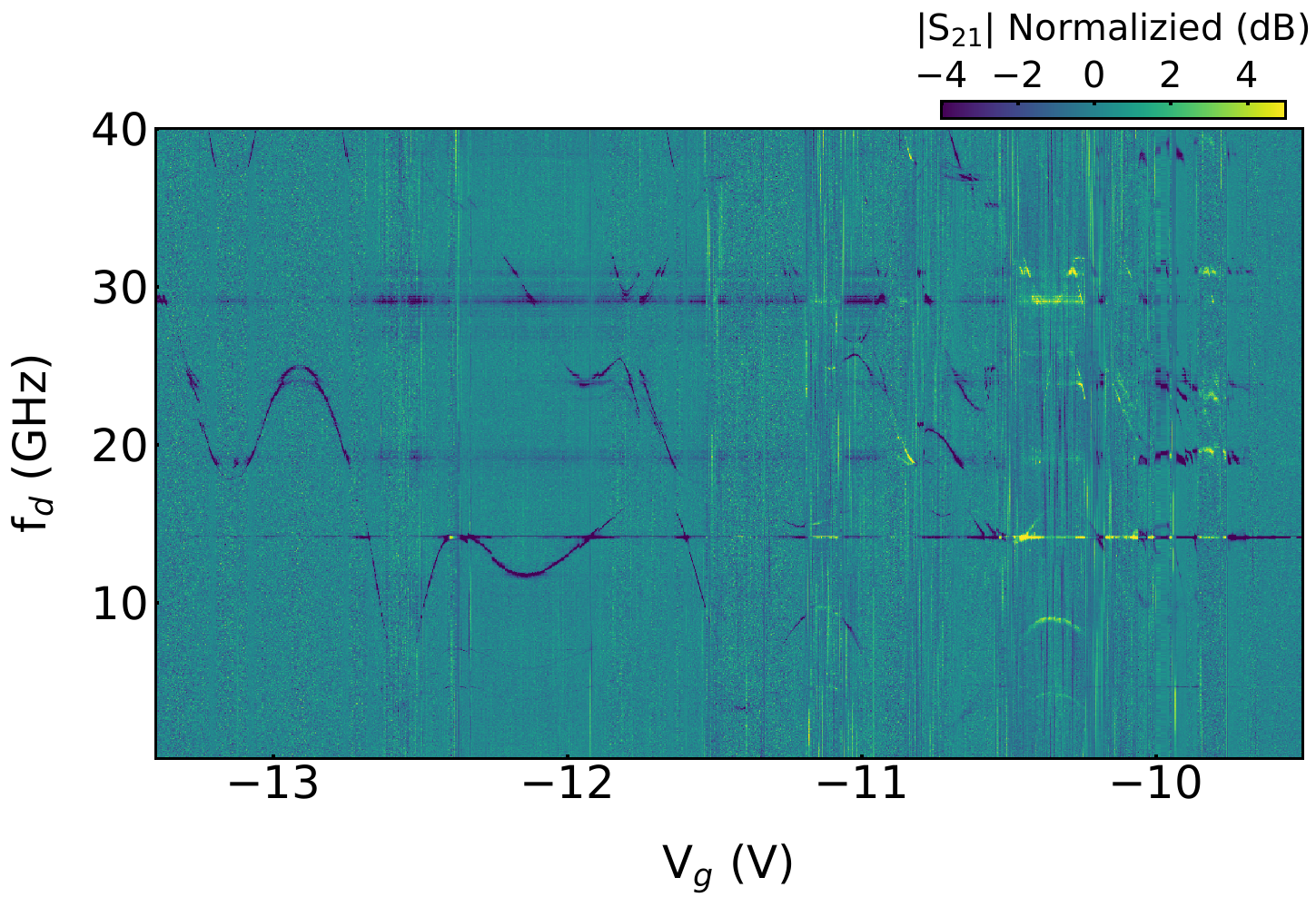}
     \caption{$|S_{21}|$ at a readout frequency $\SI{1}{\mega Hz}$ detuned from $f_{r}$, as a function of drive tone signal frequency $f_{d}$ being applied to the split gate and DC gate voltage $V_{g}$. The applied flux is fixed to $\Phi/\Phi_{0} = 0.5$. 
     }
    \label{fig:uwavespec_spec_vg}
\end{figure*}

Instead of exciting the ABS pair transitions via photons of the readout tone, we can drive the transitions directly by applying a drive tone to the split gate with frequency $f_{d}$ that capacitively couples to the ABS modes. When $f_{d} \approx f_{A}$ the ABS excitation is driven resulting in a change in $f_{r}$ and consequently the measured $|S_{21}|$ profile. By probing the resonator with a readout tone at a set frequency (slightly off $f_{r}$) and driving the transitions with a drive tone at $f_{d}$, we perform two-tone spectroscopy measurements to directly map out the dispersion of the ABS pair transitions with flux and gate voltage. The ABS pair transitions are expected to occur around $\Phi/\Phi_{0} = 0.5$ where $hf_{A}(\phi,\tau)$, where $h$ is Planck constant, is at its minimum for modes with a high enough $\tau$ to bring these transitions within the accessible range of $f_{d}$.

In \cref{fig:uwavespec_spec_sm}, we present two-tone spectroscopy measurements at different $V_{g}$ values corresponding to the presence of a single transparent mode. Approximately quadratic signal (corresponding to $f_{A}$) is observed that represents an ABS pair transitions with minimums that vary from $f_{A}\approx\SI{29.42}{\giga Hz}$ to as low as $f_{A}\approx\SI{1.80}{\giga Hz}$. Flux-independent resonances, seen as horizontal lines in the color maps, are observed corresponding to the first excited state of the resonator at $3f_{r} = \SI{14.20}{\giga Hz}$ and a fainter line corresponding to $5f_{r} = \SI{24.02}{\giga Hz}$. The presence of a single parabola for these gate voltages indicate the presence of an isolated ABS with tunable transparency confined in the constriction between the split gates. It is important to note that lower-transparency modes that lie outside our $f_{d}$ bandwidth could be present. The parabolas can be fitted to $f_{A} = 2|E_{A}|/h$, where $E_{A}$ follows \cref{eq:abs_pm} to extract $\Delta$ and $\tau$ of the ABSs. The fits and extracted values are presented as dashed lines in \cref{fig:uwavespec_spec_sm}. For the considered range of $V_{g}$, the range for $\Delta$ is seen to be $\Delta = \SI{142.2}{\micro eV}-\SI{163.1}{\micro eV}$ consistent with values for Al-InAs JJs \cite{kjaergaard_transparent_2017, mayer_superconducting_2019, hinderling_flip-chip-based_2023} and transparencies are seen to vary from $\tau = 0.833$ to near-unity transparency with $\tau = 0.9994$ extracted for $V_{g} = \SI{-11.45}{V}$. Induced Josephson resonances due to the formation of a quantum dot in the junction could play a role in the observed ultra-high mode transparency \cite{kringhoj_suppressed_2020, bargerbos_observation_2020, chidambaram2022, hinderling_flip-chip-based_2023}.

Expanding our discussion from the single-mode picture, \cref{fig:uwavespec_spec_mm}(a) considers the allowed transitions in the presence of two ABSs $\ket{0}$ and $\ket{1}$. Four allowed transitions are present: $\ket{0-}\rightarrow \ket{0+}$, $\ket{0-}\rightarrow \ket{1+}$, $\ket{1-}\rightarrow \ket{0+}$ and $\ket{1-}\rightarrow \ket{1+}$. Here, we do not consider same manifold transitions as $\ket{1-}\rightarrow \ket{0-}$ since typically these modes are occupied. The corresponding excitation spectra of $f_{A}$ with flux is presented in \cref{fig:uwavespec_spec_mm}(b) for the four allowed transitions.  

A corresponding set of two-tone spectroscopy maps is shown in \cref{fig:uwavespec_spec_mm}(c)-(i) for $V_{g}$ values that 
exhibit more than one parabola indicating the presence of multiple ABSs. For the data presented, the cases with $V_{g}<\SI{-11}{V}$ exhibit two parabolas which likely correspond to the lowest energy traces that represent $\ket{0-}\rightarrow \ket{0+}$ and $\ket{0-}\rightarrow \ket{1+}$ or $\ket{1-}\rightarrow \ket{0+}$ shown in \cref{fig:uwavespec_spec_mm}(b). Futhermore, the spectroscopy map presented for $V_{g} = \SI{-12.3}{V}$ shown in \cref{fig:uwavespec_spec_mm}(c) exhibits two prominent parabolas with minimums at $f_{d} = \SI{15.04}{\giga Hz}$ and $f_{d} = \SI{21.70}{\giga Hz}$. Remarkably, two faint parabolas with minimums at $f_{d} = \SI{21.41}{\giga Hz}$ and \SI{21.18}{\giga Hz} can also be seen. Further theoretical investigation is needed to understand the origin of these faint parabolas in close proximity to a prominent one and whether spin-splitting effects are involved. For the cases where $V_{g}>\SI{-11}{V}$, several parabolas are observed implying that the confinement is relaxed enough to allow for the presence of more than two transparent tunable modes in the constriction. It is worth mentioning that none of the additional parabolas corresponds to replicas of $f_{A}\pm f_{r}$ or higher harmonics typically seen at high drive power (see Appendix F).


To further investigate the Andreev spectrum, we perform two-tone spectroscopy while varying $V_{g}$ and setting the flux to $\Phi/\Phi_{0} = 0.5$, presented in \cref{fig:uwavespec_spec_vg}. This corresponds to tracking the minimum of the ABS pair transitions as it evolves with $V_{g}$. For $V_{g}<\SI{-13.45}{V}$, no ABS transitions are observed corresponding to the absence of transparent ABS. While the data presented in \cref{fig:uwavespec_tun} indicates the presence of a finite amount of supercurrent for $V_{g}\gtrsim\SI{-14.3}{V}$, it is likely that this finite amount of supercurrent is carried by low transparency modes characterized with an $f_{A}(\phi,\tau)$ higher than our range of $f_{d}$. At the lowest $V_{g}$ value in \cref{fig:uwavespec_spec_vg}, a single ABS transition can be resolved. As $V_{g}$ is swept in the positive direction, the confinement in the constriction is relaxed, allowing for more ABSs in the constriction and additional transitions to start emerging. Interestingly, the transition modes observed are seen to oscillate as $V_{g}$ is varied. Below $V_{g} = \SI{-12}{V}$, several ABS transitions are observed some of which are less coherently tuned with $V_{g}$. Between $V_{g} = \SI{-12}{V}$ and $V_{g} = \SI{-11.4}{V}$, a faint transition is seen to fluctuate in the few $\sim \SI{}{\giga Hz}$ regime reaching a minimum of roughly $\sim 1-\SI{2}{\giga Hz}$. Beyond $V_{g} = \SI{-10.7}{V}$, no isolated ABS transitions are observed which could be due to a dense ABS spectrum in the junction. The spectroscopy map presented in \cref{fig:uwavespec_spec_vg} represents the electrostatic tuning and mapping of several ABSs in a narrow constriction defined in a planar JJ. We note that the data presented in the form of single-tone measurements and two-tone spectroscopy exhibit discrepancies in the exact $V_{g}$ values. This is due to gate hysteresis effects and the data being taken over different cooldowns as discussed in Appendix G.


\section{Theoretical analysis} 

In the following, we discuss the observed Andreev spectra and its dependence on $V_{g}$ within a theoretical framework using tight-binding calculations. By diagonalizing the Hamiltonian described in Appendix H, the spectrum of ABSs in a constriction of a planar Al-InAs junction is calculated. In the theoretical model, the constriction and degree of confinement which is experimentally set by the voltage applied to the split gate ($V_{g}$) is represented by varying the length of the constriction $L_{con}$ and keeping the chemical potential $\mu$ constant (see Appendix H).

\begin{figure*}[ht!]
    \centering
    \includegraphics[width=1.0\textwidth]{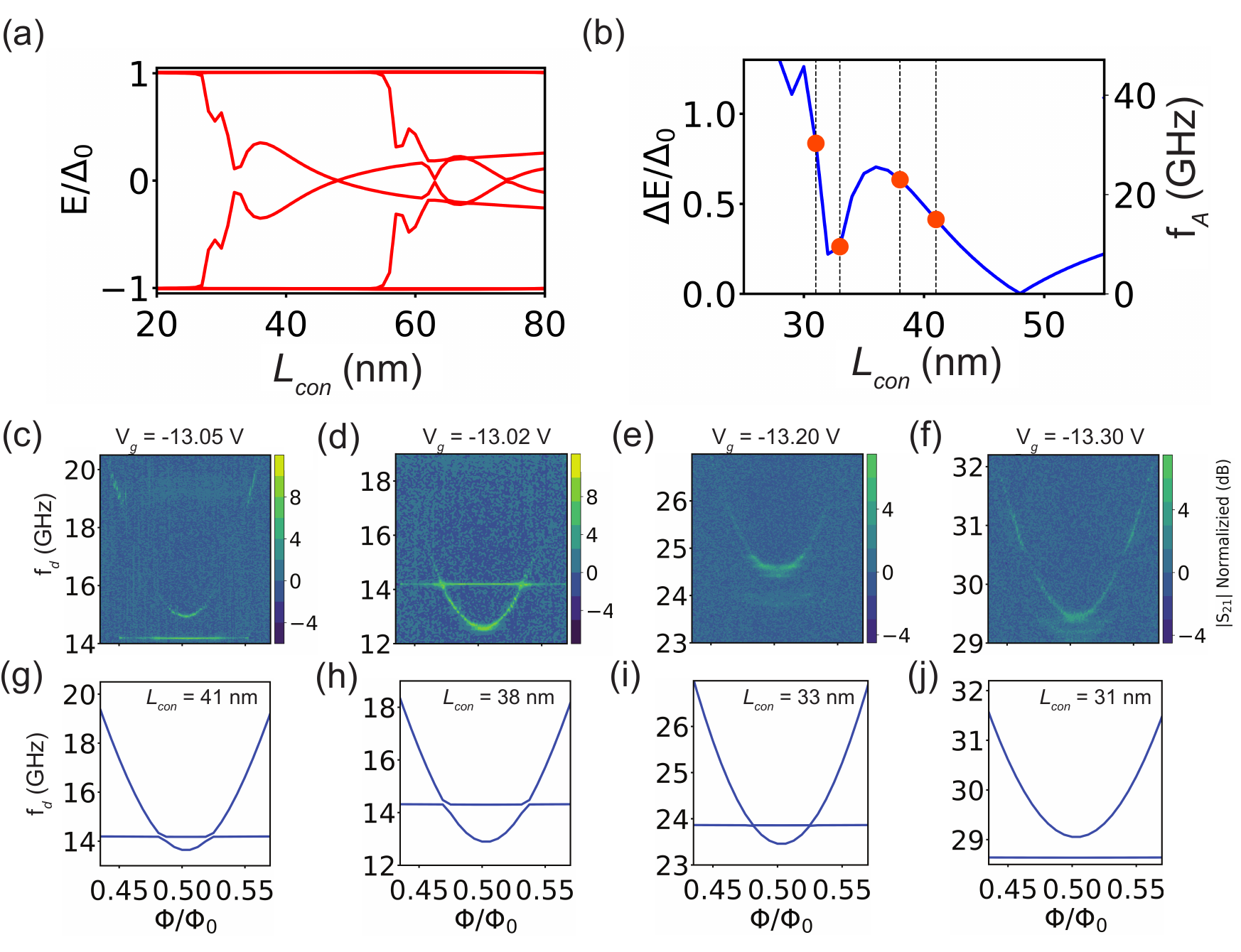}
     \caption{\textbf{(a)} Andreev spectrum as a function of the length of the constriction, $L_{con}$, at  $\phi=\pi$ in a planar junction with superconducting electrodes width $W_S=\SI{600}{\nano m}$, normal region width $W=\SI{100}{\nano m}$, and junction length $L=\SI{1}{\micro m}$ with a lattice constant of $a=\SI{2}{\nano m}$. \textbf{(b)} Transition lines for the transitions between ABS energy levels, where the orange circles mark the values of confinement, $L_{con}$, at $31, 33, 38$ and $\SI{41}{\nano m}$, which were used to calculate the results in (g)-(j).  \textbf{(c)-(f)} Two-tone spectroscopy measurements experimentally obtained as a function of $f_{d}$, and the magnetic flux $\Phi$ for different gate voltages, where (c) and (d) are re-plotted from \cref{fig:uwavespec_spec_sm}. (g)-(i) Excitation lines calculated using the Jaynes-Cummings model for $L_{con}$ at $41, 38, 33$ and $\SI{31}{\nano m}$, respectively. The parameters used for the numerical fits are $f_{r}=\SI{4.77}{\giga Hz}$, $m^{*}=0.036\,m_0$, $\Delta_0=\SI{0.15}{\milli eV}$, $\mu = \SI{12.94}{\milli eV}$ and $\alpha=\SI{10}{\milli eVnm}$.
     }
    \label{fig:theory}
\end{figure*}

The calculated Andreev spectrum as a function on $L_{con}$ is presented in \cref{fig:theory}(a) for a junction with junction length $L=\SI{1}{\micro m}$ and width $W=\SI{100}{\nano m}$. For a chemical potential $\mu = \SI{12.94}{\milli eV}$, a single subgap mode is seen to appear starting $L_{con} \approx \SI{25}{\nano m}$ and a second mode emerges at $L_{con} \approx \SI{55}{\nano m}$. In \cref{fig:theory}(b), we plot the transition lines calculated as the difference of energies between the ABS levels for the $L_{con}$ parameter space yielding a single mode (up to particle-hole symmetry and spin). The Andreev spectra and excitation spectra exhibit modes that show a non-monotonic behavior dependence on $L_{con}$ with an oscillatory pattern that is qualitatively similar to the spectroscopy results shown in \cref{fig:uwavespec_spec_vg} as a function of $V_{g}$. Further, similar to the experimental results, specific parameter regions in $L_{con}$ exhibit an high effective transparency. The size and number of these oscillations depends on the width of the junction and the superconducting leads, as well as the set $\mu$. This is mainly due to the Bohr-Sommerfeld condition for the normal-- and Andreev--reflected modes in the junction and the SC leads varying \cite{Pekerten2024b:APL}. Even for perfect transparency, the finite size of the superconducting leads causes normal modes to reenter the junction after being reflected from the boundaries of the SC leads, resulting in an imperfect effective transparency \cite{Pekerten2024b:APL}. This in turn changes the size of the oscillations in energy. Changing the normal region dimensions or changing $\mu$ affects the number of oscillations as the confinement is varied for a given number of modes, leading to a high sensitivity of the depth of the ABS energy level. Changes in $\mu$ on the order of $\Delta_0$ or smaller is seen to significantly alter the confinement size $L_{con}$ at which the ABS mode has a high effective transparency.

Finally, we consider numerically fitting the flux dispersion of the excitation (two-tone) spectra obtained experimentally by modeling the resonator-ABS coupling using a Jaynes-Cummings Hamiltonian, as described in Appendix I.
The experimental data, for specific $V_g$ values, is shown in \cref{fig:theory}(c)-(f) along with their corresponding numerical fits in \cref{fig:theory}(g)-(j) at the $L_{con}$ values marked by orange circles in \cref{fig:theory}(b). For a spin-orbit coupling strength $\alpha=\SI{10}{\milli eVnm}$, the fits are seen to provide a good approximation to the experimental data. We note that the fits are sensitive to the value of $\mu$ as discussed earlier (see Appendix J).

\section{Conclusion}

In summary, by embedding a wide planar Al-InAs Jospheson junction into a superconducting circuit, we probe ABSs in a electrostatically-defined narrow constriction in the junction. We observe evidence of resonator-ABS coupling in the form of avoided crossings with resonant frequency of the resonator due to virtual photon exchange between the resonator and the ABSs. Further, we directly drive ABS pair transitions by applying a drive tone to the split-gate and observe single and multiple ABS pair transitions and explore the tunability of these ABS pair transitions with $V_{g}$ to observe a rich, complex spectra exhibiting the presence of multiple tunable ABS transitions. Finally, we support the experimental results with tight-binding simulations that align with the experimental data. These results open the door to using microwave spectroscopy techniques to characterize the Andreev spectrum of wide planar Josephson junctions fabricated on superconductor-semiconductor heterostructures for applications in Andreev spin qubits and induced topological superconductivity.

\section*{Acknowledgements}

We thank Max Hays, Valla Fatemi, Peter Schuffelgen and Christian Dickel for fruitful discussions. We acknowledge support from MURI ONR award no. N00014-22-1-2764. NYU team also acknowledge support from the Army Research Office agreement W911NF2110303 and W911NF2210048. W.M.S. acknowledges funding from the ARO/LPS QuaCR Graduate Fellowship. The UB team was partially supported by NSF ECCS-2130845. Computational resources were provided by the University at Buffalo Center for Computational Research.

\

\section*{Appendix A: Design and device fabrication}

\begin{figure}[htp]
    \centering
    \includegraphics[width=0.45\textwidth]{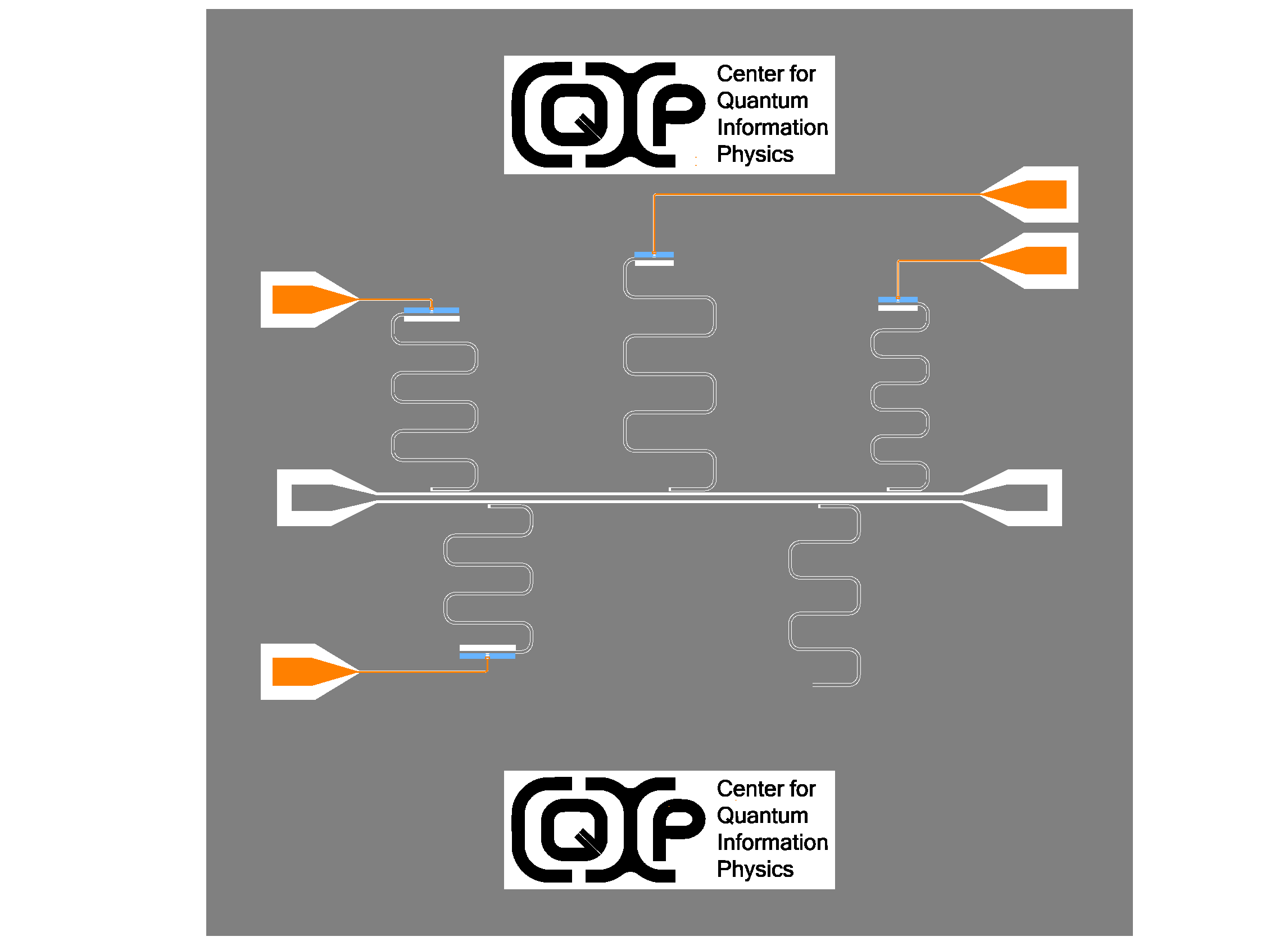}
    \caption{Schematic of the chip design with $\lambda/4$ CPWs four of which are inductively coupled to superconducting loops with an Al-InAs Josephson junction and one is a bare resonator.} 
    \label{fig:chip_design}
\end{figure}

\begin{figure}[htp]
    \centering
    \includegraphics[width=0.45\textwidth]{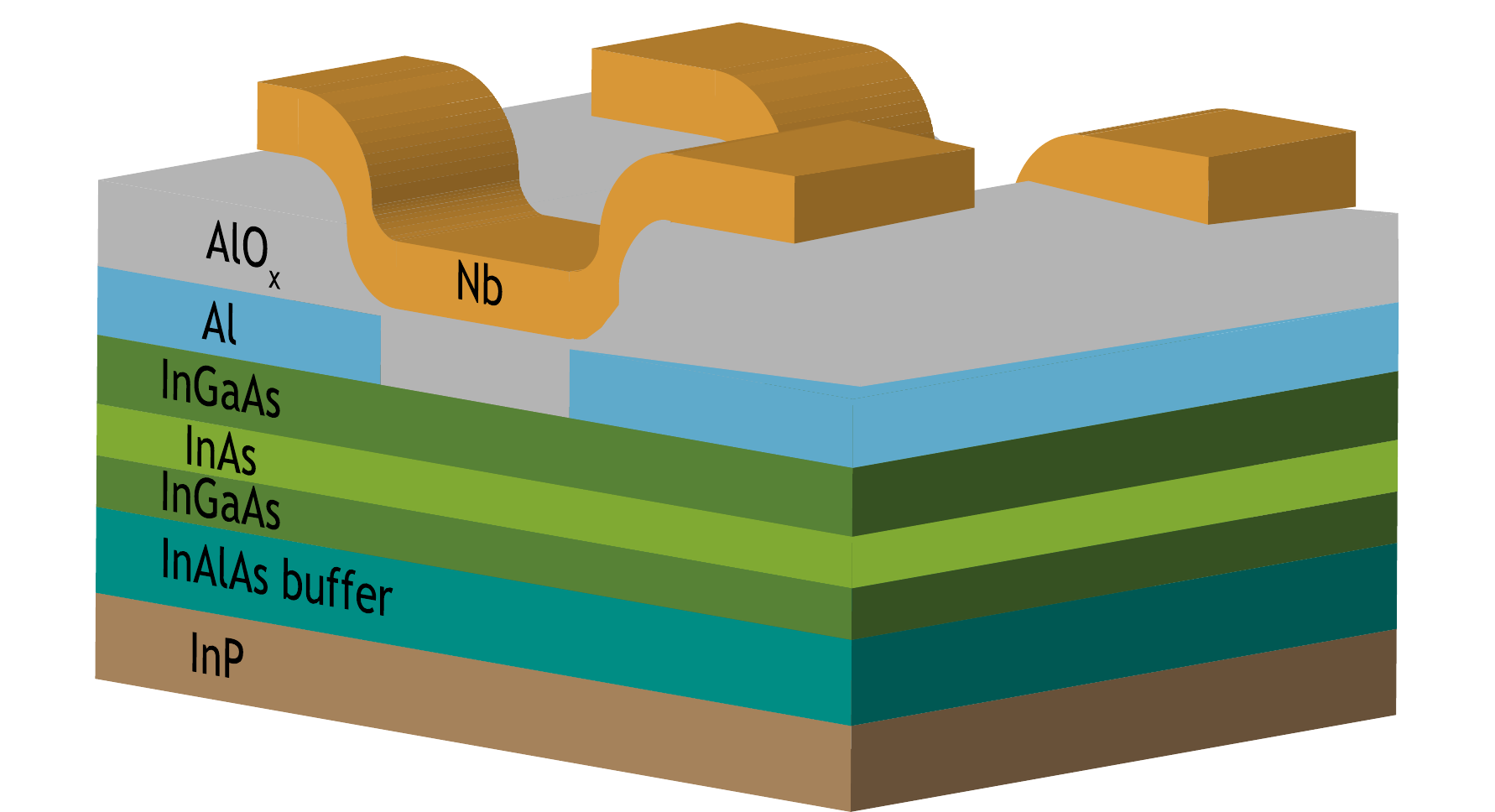}
    \caption{Schematic of the material heterostructure with a junction, made of Al superconducting contacts and an InAs surface quantum well. A layer of AlO$_\text{x}$ is deposited as a gate dielectric followed by a patterned split gate made of Nb.} 
    \label{fig:hetero}
\end{figure}

The design of the microwave circuit was constructed using Qiskit Metal \cite{Qiskit_Metal} and rendered in Ansys's high frequency simulation software (HFSS) \cite{ansys} to simulate the expected resonant frequency, external quality factors, and electromagnetic field distribution. The chip design consists of hanger $\lambda/4$ CPW resonators coupled to a common transmission feedline. The external quality factor is designed to be $Q_\mathrm{ext}\sim 9600$. The design includes four resonators (one of which is studied in this work) that are inductively coupled superconducting loops with JJs and one ``bare'' resonator that is not inductively coupled to a loop that is used as a reference, as shown in \cref{fig:chip_design}.

The superconducting loop and Josephson junction devices are fabricated on an InAs near-surface quantum well grown by molecular beam epitaxy on a \SI{500}{\micro m} thick InP substrate. After thermal oxide desorption, an In$_{x}$Al$_{1-x}$As graded buffer layer is grown to reduce strain on the InAs active region, where the composition $x$ is graded from 0.52 to 0.81. The InAs 2DEG is confined between In$_{0.81}$Ga$_{0.19}$As top and bottom barriers. Finally, a $\sim\SI{25}{\nano m}$ thick film of Al is deposited \textit{in-situ}. The detailed growth procedure of the III-V heterostructure is outlined in Refs.\cite{Shabani2016, Kaushini2018, strickland2022}.

The device is fabricated through a series of electron beam lithography steps using spin-coated PMMA resist. First, we define the superconducting loop and an area for the ground plane and microwave circuit by chemically etching the Al using Transene Al etchant type-D and the III-V layers using an III-V etchant consisting of phosphoric acid (H$_3$PO$_4$, 85\%), hydrogen peroxide (H$_2$O$_2$, 30\%) and deionized water in a volumetric ratio of 1:1:40. The planar junctions in the loop are defined to be $\sim\SI{5}{\micro m}$ long. The junction gap is then defined by etching a $\sim\SI{100}{\nano m}$ strip of Al. Considering  electron mean free path of $\sim\SI{420}{\nano m}$ as measured by low-temperature Hall measurements, the junction should be in the short ballistic regime. The microwave circuit is then defined through a process of patterning with electron beam lithography, sputtering $\SI{100}{\nano m}$ of Nb and a liftoff process. This is followed by the deposition of a blanket layer of \SI{40}{\nano m} layer of AlO$_\text{x}$ at $\SI{40}{C^\circ}$ as a gate dielectric by atomic layer deposition, followed by a sputtered split gate made of $\SI{25}{\nano m}$ Nb layer using liftoff. The distance between the split gates over the junction is defined to be $\sim \SI{100}{\nano m}$. To ensure the gate climbs over the mesa wall, $\SI{700}{\nano m}$ of Nb is sputtered in the region where the gate climbs over the mesa wall in a separate lithography and deposition step. A schematic of the junction heterostructure after fabrication is shown in \cref{fig:hetero}.

\section*{Appendix B: Measurement setup}

\begin{figure*}[htpb]
    \centering
    \includegraphics[width=1.0\textwidth]{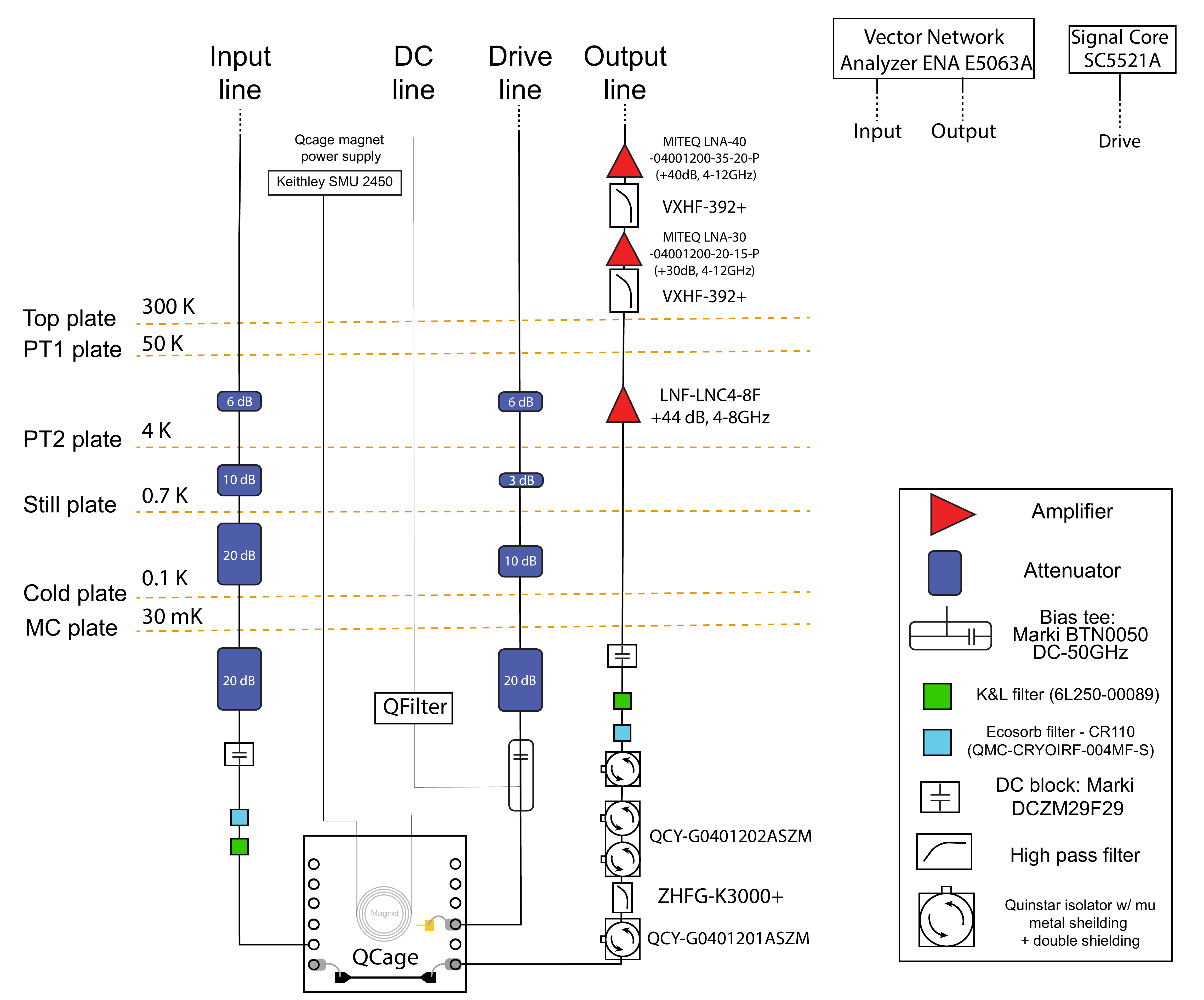}
    \caption{Schematic of the cryogenic and room temperature measurement setup.}
    \label{fig:fridge}
\end{figure*}

A schematic of the cryogenic and room temperature measurement setup is shown in \cref{fig:fridge}. Measurements are conducted in an Oxford Triton dilution refrigerator. The sample is placed in a QCage, a microwave sample holder, and connected to the printed circuit board by aluminum wirebonds. Probe signals are sent from a vector network analyzer (VNA) or an AWG and attenuated by \SI{-56}{dB} with attenuation at each plate as noted. The signal then passes through a 1-\SI{10}{\giga Hz} bandpass filters in the form of Eccosorb and K\&L filters. The signal is sent through the sample, returned through a series of isolators and bandpass filters. The signal is then amplified with a low noise amplifier mounted to the 4K plate and two room temperature amplifiers (MITEQ) outside the fridge. The output signal is then measured with a VNA.

\section*{Appendix C: Readout power dependence}

\begin{figure*}[ht!]
    \centering
    \includegraphics[width=1.0\textwidth]{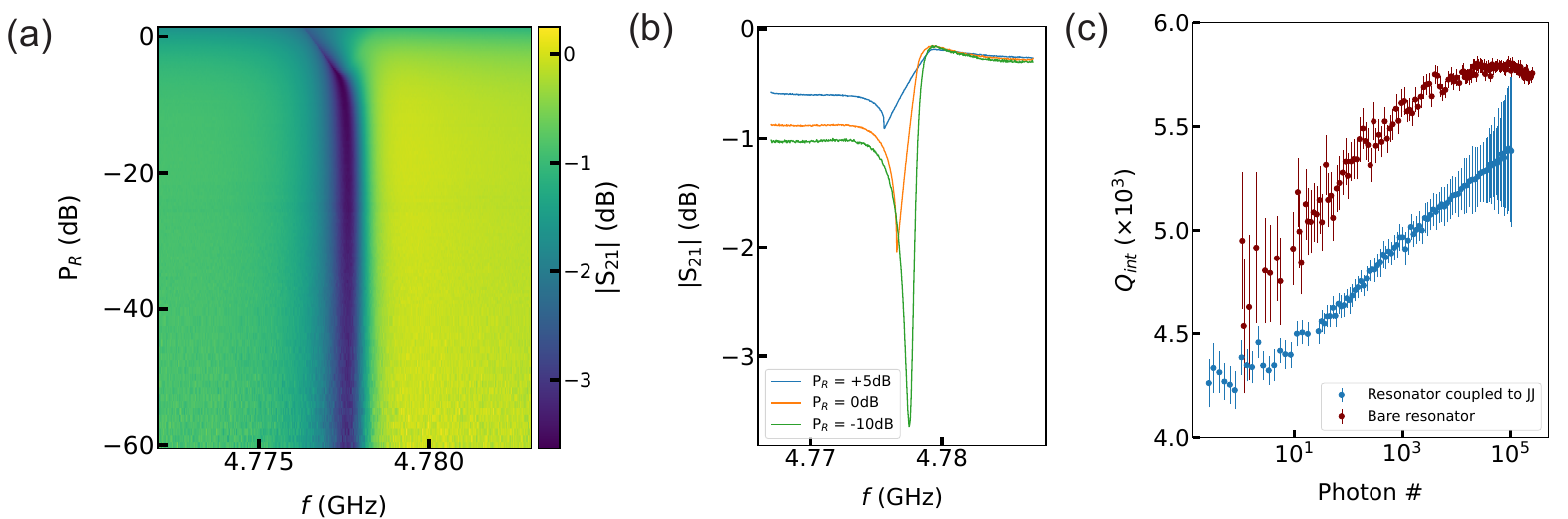}
     \caption{ \textbf{(a)} Power-dependent response of the resonator represented as the magnitude of transmission coefficient $|S_{21}|$ as a function of readout frequency ($f$) and change in readout power $P_{R}$. \textbf{(b)} Linecuts from the colormap corresponding to different values of $P_{R}$. \textbf{(c)} Dependence of extracted internal quality factor $Q_{int}$ on function $P_{R}$. For the resonator coupled to a JJ, these measurements are done at $V_{g} = \SI{0}{V}$ and with no applied flux.
     }
    \label{fig:uwavespec_power_dep}
\end{figure*}

The readout power-dependent response of the resonator shown in \cref{fig:uwavespec_power_dep}. At low power, the resonant frequency of the resonator is measured to be $f_{r} = \SI{4.777}{\giga Hz}$. As the power increases, $f_{r}$ shifts towards lower values as seen in \cref{fig:uwavespec_power_dep}(a) and a discontinuity appears in the resonance shape as seen in the linecuts of $|S_{21}|$ shown in \cref{fig:uwavespec_power_dep}(b). The observed high-power response is due to the nonlinearity of the junction which results in the response becoming multivalued with two metastable solutions existing at a single frequency; a process referred to a bifurcation \cite{yurke1989, HoEom2012, siddiqi2004, vijay2009, phan2022}. The nonlinearity and bifurfaction effects due to the coupling of an Al-InAs junction to a CPW resonator is discussed further in Ref. \cite{strickland_superconducting_2023}.

Extracting the internal quality factor $Q_{int}$ from the linecuts of \cref{fig:uwavespec_power_dep}(a) where the resonance lineshape is continuous and Lorentzian, we plot the power dependence, expressed in  terms of the number of photons, of $Q_{int}$ in \cref{fig:uwavespec_power_dep}(c). A downward trend is observed for $Q_{int}$ with photon number, with a decrease of about $\%20$ from $\sim10^{5}$ photons to the single photon regime. Comparing this to the power-dependence of $Q_{int}$ for a bare resonator (resonator without a coupled loop) on the same chip, we observe a similar downward trend in $Q_{int}$ at lower photon number. The dependence of $Q_{int}$ on photon number is consistent with the presence of two-level systems (TLSs) in our system \cite{wenner_surface_2011, sage_study_2011, zmuidzinas_superconducting_2012}. Since this is the case for both the bare resonator and the resonator coupled to a JJ, we can conclude that these TLS loss are inherent to the resonators.

\section*{Appendix D: Gate dependence of $Q_{int}$}

\begin{figure}[ht!]
    \centering
    \includegraphics[width=0.45\textwidth]{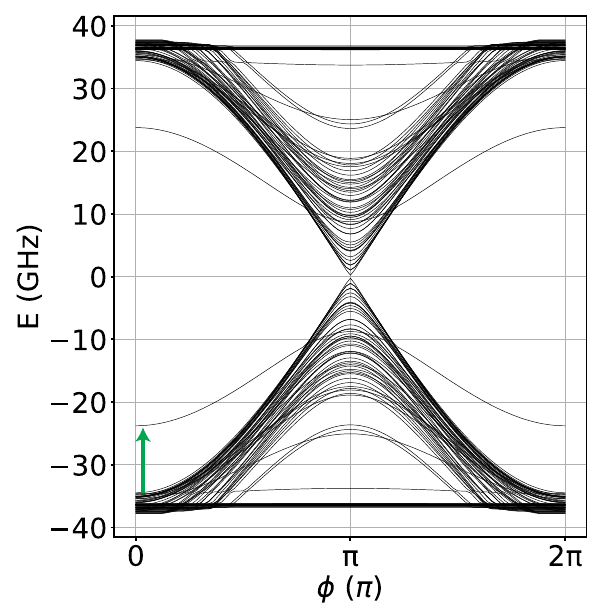}
     \caption{Calculated energy spectrum of the Andreev bound states in a wide Al-InAs junction. The results obtained are for a JJ with length $L = \SI{1}{\micro m}$, normal region length $W = \SI{100}{\nano m}$, length of the superconductor $W_{S} = \SI{1.5}{\micro m}$, superconducting gap $\Delta = \SI{150}{\micro eV}$, carrier density $n = 4 \times 10^{11}\SI{}{\centi m^{-2}}$ and effective electron mass $m^* = 0.04m_{e}$ where $m_{e}$ is the electron mass. Green arrow represents a transition between a short junction mode typically pinned to the continuum and a long junction separated from the continuum at $\phi = 0$. Details of the simulation of the Andreev spectrum is provided in Ref.\cite{elfeky_quasiparticle_2023}.
     }
    \label{fig:abs}
\end{figure}

The dependence of $Q_{int}$ on $V_{g}$ is presented in the inset of \cref{fig:uwavespec_tun}(a). In the $V_{g}$ range corresponding to the modes directly under the junction being suppressed and the constriction becoming defined, $Q_{int}$ is seen to drop. A possible explanation for this behavior is that in this range of $V_{g}$ the spectrum of Andreev bound states evolves in a way that results in the energy spacing at $\phi = 0$ between specific modes that were previously largely separated to become closer making transitions more likely. This could be the case for the energy spacing between a short-junction mode and a long-junction mode, which are separated from the continuum at $\phi = 0$ by an energy $\delta$ that diminishes as the junction length decreases \cite{dartiailh_missing_2021, elfeky_evolution_2023, elfeky_quasiparticle_2023}. While at zero temperature all the states corresponding to the negative branches are occupied and no transitions are possible, at finite temperature the occupation of ABSs follows a Fermi-Dirac distribution which means that a finite amount of negative-energy modes can be unoccupied, allowing for these transitions to occur between the modes. As the constriction becomes defined and the effective junction width narrows down, $\delta$ is expected to be suppressed, thus excitations between these modes mediated via the photons of the readout tone at $f_{r}$ become more favorable. In that case, multi-photon processes could play a role in these excitations. 
Since the VNA measurements here are done on a relatively long time-scale (IF bandwidth of 100-\SI{1000}{Hz}), with respect to the timescale of these excitation and relaxation processes, the effect of these excitations are averaged out and could be interpreted as noise and dephasing in the resonant frequency of the resonator reflected as a decrease in $Q_{int}$ due to these mode-to-mode excitations. After the junction is narrow enough or when the constriction is defined, the long-junction modes are completely suppressed and $Q_{int}$ increases again and plateaus back to a constant value. Further theoretical investigation is required to understand the origin of the $Q_{int}$-dependence on $V_{g}$.

\section*{Appendix E: Andreev Transitions through multi-photon processes}

\begin{figure}[ht!]
    \centering
    \includegraphics[width=0.45\textwidth]{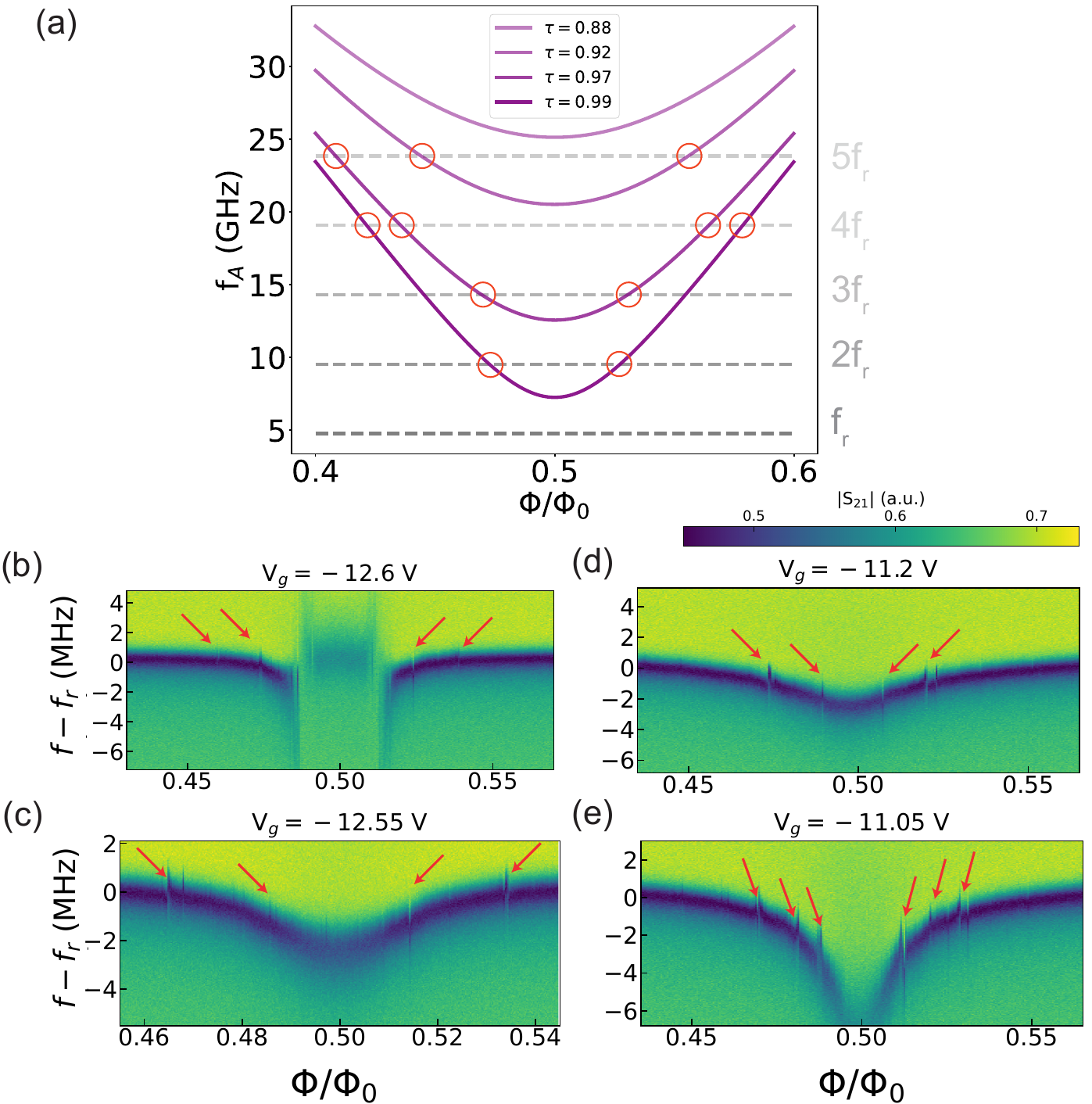}
     \caption{\textbf{(a)} The excitation energy $f_{A} = 2|E_{A}| = 2\Delta \sqrt{1-\tau\sin^2(\phi/2)}$ corresponding to a transition between the negative and positive branches of an ABS for ABSs with transparencies $\tau$ that correspond to $f_{A}>f_{r}$ for all values of $\Phi/\Phi_{0}$. Here, the induced superconducting gap is assumed to be $\Delta = \SI{150}{\micro eV}$. Grey dashed lines correspond to the frequency of the ground state of the resonator ($f_{r}$) and multi-photon states of the resonator. Orange circles mark the crossing of $f_{A}$ with photon states of the resonator. \textbf{(b)-(e)} Transmission coefficient $|S_{21}|$ as a function of the readout frequency and applied flux $\Phi$ around $\Phi/\Phi_{0} = 0.5$ for different $V_{g}$ applied to the split gate. Orange arrows correspond faint peaks observed in $f_{r}$.
     }
    \label{fig:multi-photon}
\end{figure}

The data presented in \cref{fig:uwavespec_ac} is taken with a readout power corresponding to the few photon regime (an average of $\approx 1.2-4.6$ photons depending on the exact value of flux). For specific values of $V_{g}$, the modes are not transparent enough for $f_{A}$ to cross $f_{r}$ and the observation of a prominent avoided crossing. In that case, multi-photon processes corresponding to 2, 3 or more photons are possible which can cause some weak coupling to the ground state of the resonator. Consequently, when $f_{A}$ crosses multiples of $f_{r}$, represented in \cref{fig:multi-photon}(a) as orange circles, a faint avoided crossing (peaks/dips) can be expected. In \cref{fig:multi-photon}(b)-(e), we plot the flux dispersion for different $V_{g}$ values where peaks/dips that are symmetric about half-flux are seen, pointed out by orange arrows.

In \cref{fig:multi-photon}(b)-(e), more than one peaks is seen on each side of flux which either corresponds an ABS with $f_{A}$ that crosses different multi-photon resonator states (i.e. $2f_{r}$, $3f_{r}$ etc.) or more than one ABS with different $f_{A}$ that cross one of these multi-photon resonator states. Given the low readout photon number, we expect that multi-photon processes corresponding to four photons and higher to not contribute significantly in these processes. If we consider the outer peaks seen for $V_{g} = \SI{-11.2}{V}$ and assume that they are due to two-photon (three-photon) processes at $2f_{r}$ ($3f_{r}$), this would correspond to a mode with $\tau \approx 0.9948$ ($\approx 0.9729$), setting $\Delta = \SI{150}{\micro eV}$. For $V_{g} = \SI{-12.6}{V}$, in addition to the prominent avoided crossings, the dispersion exhibits two fainter peaks on each side at $\Phi/\Phi_{0} \approx 0.4607$/0.5386, and at $\Phi/\Phi_{0} \approx 0.4741$/0.5238. Again here, the peaks could be due to ABS-resonator coupling through higher multi-photon processes if the mode that resulted in the prominent avoided crossing or another different mode. Assuming the prominent avoided crossing at $\Phi/\Phi_{0} \approx 0.485$ is due to a single-photon process, which corresponds to a $\approx0.9977$, the same mode is expected to cross $2f_{r}$ at $\Phi/\Phi_{0} = 0.46085$ which likely corresponds to outermost peak. The innermost peak could correspond to the presence of additional less-transparent ABSs that are shallower and so weakly couple to the resonator through multi-photon processes. 

\section*{Appendix F: High drive power replicas}

For the transitions observed in two-tone measurements, typically at high drive powers, for a given $f_{A}$ replicas corresponding to $f_{A}\pm f_{r}$ can appear due to an interference of photons from the readout tone and photons from the drive tone \cite{conner_superconducting_2021, wesdorp2022}. Replicas corresponding to $f_{A}/n$ can also be expected at high drive power. For the two-tone experiments presented in this work (\cref{fig:uwavespec_spec_sm}, \cref{fig:uwavespec_spec_mm}, \cref{fig:uwavespec_spec_vg}), the power of the drive tone is set to be low enough such that these replicas do not appear. Further, to keep the effective drive power roughly constant, the room temperature attenuation on the drive line is varied with the drive frequency, to account for the additional attenuation of the drive signal in the fridge RF lines at high frequencies.

To confirm that the additional parabolas seen in \cref{fig:uwavespec_spec_mm} are not due to trivial replicas, let's consider the $V_{g} = \SI{-12.2}{V}$ case. As seen in \cref{fig:uwavespec_spec_mm}(e), the minimum of the bottom parabola occurs at $f_{d} = \SI{11.87}{\giga Hz}$. At high enough drive power, one can expect to see a replica corresponding to $f_{d}+f_{r} = \SI{16.64}{\giga Hz}$. However, at this low drive power, the second parabola is observed with a minimum at $f_{d} = \SI{17.51}{\giga Hz}$ which implies that this parabola corresponds to the presence of a second ABS pair transition in the constriction consistent with the picture presented in \cref{fig:uwavespec_spec_mm}(a)-(b).

\section*{Appendix G: Correlating single-tone and two-tone measurements}

\begin{figure*}[ht!]
    \centering
    \includegraphics[width=1.0\textwidth]{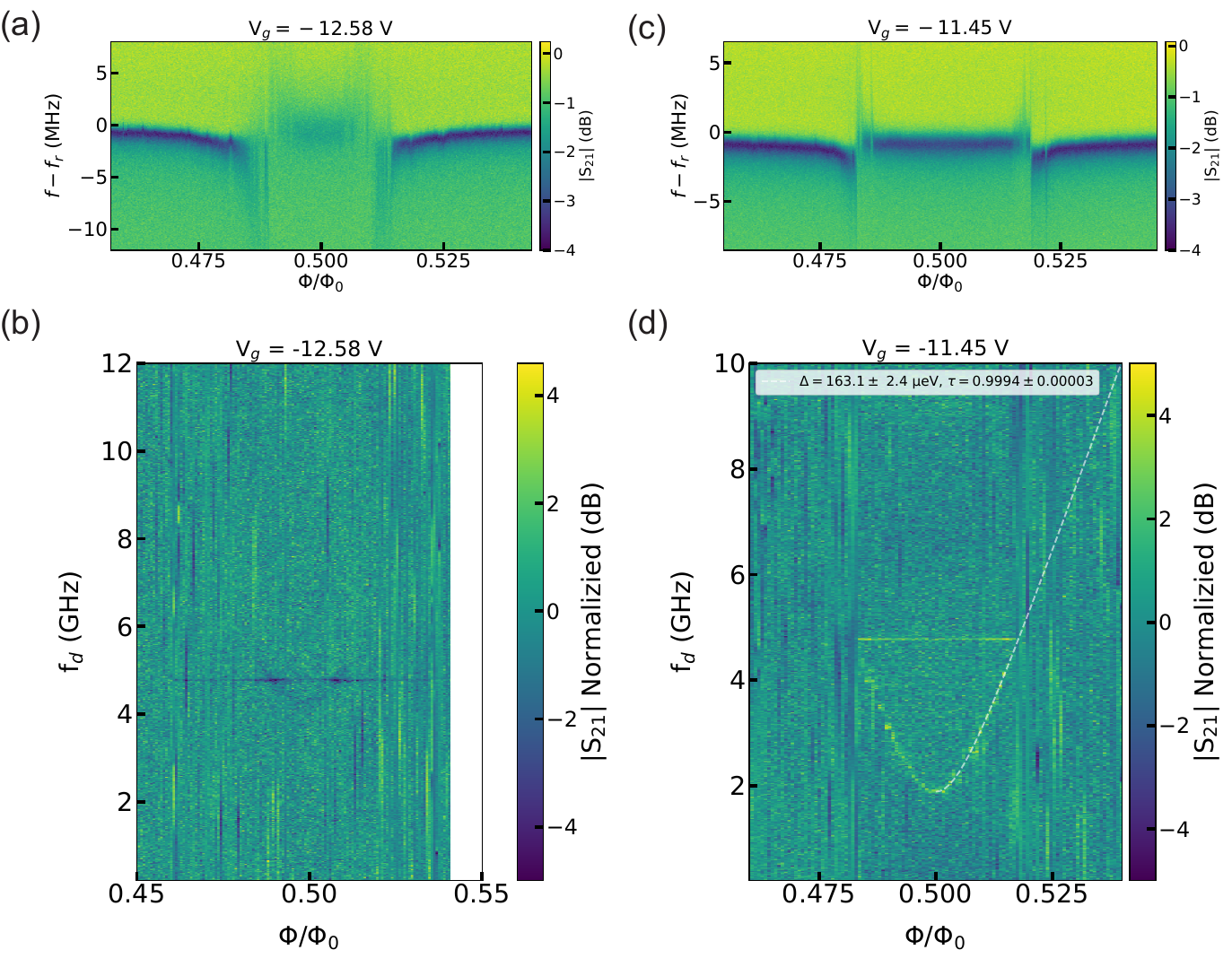}
     \caption{\textbf{(a)}, \textbf{(c)} Single-tone and \textbf{(b)}, \textbf{(d)} two-tone measurements performed sequentially to each other without the value of $V_{g}$ being changed between the two measurements. This is done to make sure gate hysteresis effects do not play a factor and that the same effective $V_{g}$ is applied in each case.
     }
    \label{fig:uwavespec_single-two}
\end{figure*}

The data presented in the form of single-tone (readout tone) measurements and two-tone (readout tone plus drive tone) spectroscopy in this work exhibit discrepancies in the exact $V_{g}$ values used. This is due to gate hysteresis effects and the data being taken over different cooldowns which changes the ``effective'' $V_{g}$. Each cooldown results in a different disorder configuration and consequently a slight shift in the gate voltage and corresponding chemical potential parameter space. In \cref{fig:uwavespec_single-two}, we present single-tone and two-tone measurements that were performed at the same $V_{g}$ without $V_{g}$ being changed between the single-tone and two-tone measurements, to ensure gate hysteresis effects do not play a factor. For the two $V_{g}$ values shown, we do in fact see a clear correlation between the position of the avoided crossings in the single-tone measurement and an faint ABS pair transition crossing $f_{r}$ in the two-tone measurement.

It is important to note that in the case of single-tone measurements, the ABS pair excitations are driven through phase fluctuations caused by flux fluctuations from the readout resonator (inductive coupling). On the other hand, the two-tone measurements drive the ABS pair excitations through fluctuations in the chemical potential caused by the drive tone applied to the gate (capacitive coupling). The coupling to different transitions can be different depending on the type of drive being used. In the two-tone spectroscopy map seen in \cref{fig:uwavespec_spec_vg}, we see that different gate and drive frequency parameter spaces exhibits transitions that vary in prominence. This implies a variation in the coupling of these transitions to the drive tone or a probabilistic nature to the occurrence of such transitions. This could be a result of several factors other than the detuning of the transition frequency to the resonant frequency of the resonator. The coupling of the ABS transitions and drive tone could be gate-dependent where for specific values of $V_{g}$, the chemical potential is more or less sensitive to the perturbations caused by the drive tone. The geometry of the split gate might also allow for inhomogeneous and non-trivial coupling to transitions in the junction.

\section*{Appendix H: Tight-binding Calculations}

In the theoretical analysis presented in this work, we utilize the Kwant package \cite{Groth2014:NJP} for our discretized tight-binding (TB) numerical calculations describing the JJ dynamics. The TB form of the Bogoliubov-de Gennes (BdG) Hamiltonian describing the planar JJ is 
\begin{align}
H_0^\mathrm{TB}	&= \sum_{i, j} H_0^\mathrm{onsite}\,\ket{\mathbf{r}_{i,j}}\bra{\mathbf{r}_{i,j}}  +\big(H_0^\mathrm{up}\,\ket{\mathbf{r}_{i,j+1}}\bra{\mathbf{r}_{i,j}} \nonumber\\
		         	& \quad  + H_0^\mathrm{right}\,\ket{\mathbf{r}_{i+1,j}}\bra{\mathbf{r}_{i,j}} + \mathrm{h.c.}\big),
\label{eq:SM:TBHamiltonian}
\end{align}
where h.c. represents a Hermitian conjugate and
\begin{align}
H_0^\mathrm{onsite}	&= [4t -\mu^\mathrm{TB}(\mathbf{r}_{i,j})]\,\tau_z + B^\mathrm{TB}\,\sigma_y \nonumber\\
					&+ \Delta(\mathbf{r}_{i,j}) \, \left(e^{i\,\phi (\mathbf{r}_{i,j})/2} \tau_+ + e^{-i\,\phi(\mathbf{r}_{i,j})/2} \tau_- \right)\, ,  \nonumber\\
H_0^\mathrm{up}		&= -t \,\tau_z + \frac{\alpha^\mathrm{TB}}{2a}\,\sigma_x\tau_z\, , \nonumber\\
H_0^\mathrm{right}	&= -t \,\tau_z - \frac{\alpha^\mathrm{TB}}{2a}\,\sigma_y\tau_z\, .
\label{eq:SM:TBHamiltonian_Parts}
\end{align}
In \cref{eq:SM:TBHamiltonian_Parts}, $t=\hbar^2/(2 m^\ast a^2)$ is the hopping parameter, $a$ is the TB lattice parameter and $m^\ast$ is the effective mass. The indices $i$ and $j$ describe the discretized $x$- and $y$-coordinates of the TB lattice. We consider a uniform chemical potential $\mu^\mathrm{TB}(\mathbf{r}_{i,j}) = \mu(\mathbf{r}) = \mu$ and a uniform Rashba spin-orbit coupling (SOC) strength $\alpha^\mathrm{TB} = \alpha$. The applied magnetic field,  $B^\mathrm{TB}=B$ is zero. $\sigma_i$ ($\tau_i$) are Pauli (Nambu) matrices in the spin (particle-hole) space with $\tau_\pm=(\tau_x\pm i\tau_y)/2.$ $\Delta(\mathbf{r}_{i,j}) = \Delta_0 \,\Theta(|i-L/2a|)$ is the pair potential and is nonzero and uniform in the superconducting (S) regions. $\phi(\mathbf{r}_{i,j}) = \phi_0\, \mathrm{sgn}(i) \,\Theta(|i-W_N/2a|)$ is the superconducting phase difference across the junction, where $\Theta$ is the step function. 

An applied microwave drive is described as a perturbation that couples to and effectively leads to a change in the phase difference $\phi\rightarrow \phi_0 + \delta \phi(t)$ \cite{Desposito2001:PRB, Zazunov2003:PRL, Janvier2015:S, Hays2021:S, Olivares2014:PRB, Vayrynen2015:PRB, Pekerten2024c:PRB}. The TB form of this perturbation is given by
\begin{align}\label{eq:SM:TB_MWHamiltonian}
H^\mathrm{TB}_\mathrm{MW}	&= \frac{i}{2} \sum_{i, j} \Delta(\mathbf{r}_{i,j}) \, \mathrm{sgn}(i)\,\ket{\mathbf{r}_{i,j}}\bra{\mathbf{r}_{i,j}} \nonumber\\
							& \times \, \left( e^{i\,\phi(\mathbf{r}_{i,j})/2} \tau_+ -  e^{i\,\phi(\mathbf{r}_{i,j})/2} \tau_- \right).
\end{align}
The energy spectra $E_n$ of the ABS in the junction and the matrix elements $\mathcal{M}_{mn}$ are calculated using $H^\mathrm{TB} = H_0^\mathrm{TB} + H^\mathrm{TB}_\mathrm{MW}$ as described by \cref{eq:SM:TBHamiltonian} and \cref{eq:SM:TB_MWHamiltonian}.

\begin{figure*}[tbh]
\centering
\includegraphics[width=\textwidth]{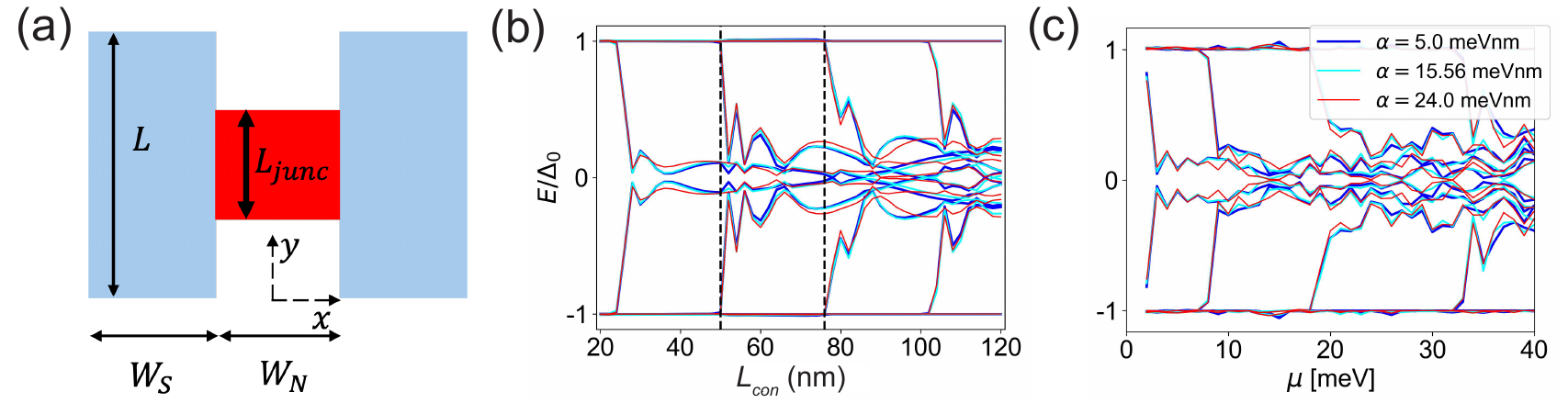}
\caption{(a) Schematic of a planar JJ with length $L$ and a constriction of length $L_{con}$. $W_S$ ($W_N$) is denoted as the width of the superconducting (normal) region. (b) The ABS spectrum of the junction as a function of $L_{con}$ for $\mu=\SI{20}{\milli eV}$ and for different $\alpha$ values. Dashed vertical lines represent the value at which more modes enter the spectrum. (c) The ABS spectrum of the junction as a function of $\mu$ for $L_{con}=\SI{80}{\nano m}$ and for different $\alpha$ values indicated in the legend. For both (b) and (c), $B=\SI{0}{T}$, $W_S=\SI{600}{\nano m}$, $W_N=\SI{100}{\nano m}$, $L=\SI{1}{\micro m}$, $\Delta_0=\SI{0.22}{\milli eV}$, $T=\SI{20}{\milli K}$, $T_c=\SI{1.4}{K}$, $B_c=\SI{250}{\milli T}$ and $a=\SI{2}{\nano m}$.}
\label{FIG:sim_schematic} 
\end{figure*}

A schematic of the simulated planar JJ with the constriction is presented in \cref{FIG:sim_schematic} (a). The confinement of the normal region (shown in red in \cref{FIG:sim_schematic} (a)) along the $y$-direction is experimentally governed by the voltage applied to the split-gate. In the theoretical model, the degree of confinement can be represented either by changes in constriction length $L_{con}$ or the chemical potential $\mu$. As the confinement $L_{con}$ is relaxed for a given $\mu$, or as $\mu$ is increased for a given $L_{con}$, more modes enter the gap as seen in \cref{FIG:sim_schematic}(b) and (c). We note that changing $\mu$ of the system for a given $L_{con}$ is qualitatively equivalent to changing $L_{con}$ for a given $\mu$. This is evident by comparing \cref{FIG:sim_schematic}(b) and (c) where similar dependence on each parameter is observed. For the theoretical analysis discussed in the main text, we set $\mu$ to a constant value and vary $L_{con}$.

\section*{Appendix I: Resonator-ABS Modeling}

To model the coupling between the microwave resonator and the ABS in the junction we use the Jaynes-Cummings Hamiltonian \cite{koch_charge-insensitive_2007, park_adiabatic_2020}:
\begin{equation}\label{EQN:JC}
    H_{JC} = H_{0} + H_{r} + H_{int},
\end{equation}
where $H_0 = \sum_{i,j}\ket{\varphi_{i}}\bra{\varphi_{i}}H_{0}^{TB} \ket{\varphi_{j}} \bra{\varphi_{j}}$ is the TB Hamiltonian for the Josephson junction in the basis of its eigenvectors $\left\{ \ket{\varphi_{i}} \right\}$, $H_{r} = \hbar\omega_{r}(aa^{\dagger} + 1/2)$ is the resonator Hamiltonian with $\omega_R$ being the resonance frequency of the resonator, and $H_{int} = M\, I_{r} \, I_\mathrm{JJ}$ describes the interaction between the resonator and the SQUID-loop. Here, $I_{r} = (1/L)\,\sqrt{\hbar\,/\,2\omega_r\,C}\,(a+a^{\dagger})$ with $L$ ($C$) being the effective inductance (capacitance) of the resonator \cite{Park2017:PRB} and $I_\mathrm{JJ} = (2e/\hbar)\, \partial H_0/\partial \phi$ is the perturbation introduced by the microwave drive of the superconducting phase difference $\phi$ across the junction. 

Denoting the number of photon modes in the resonator as $n$, and its corresponding state $\ket{n}$, the basis of the system is formed by $\{\ket{\varphi_{i},n}\}$. Thus, the matrix elements of the Hamiltonian in \cref{EQN:JC} are
\begin{align}\label{EQN:JCMatrixElements}
        & \bra{\varphi_{j},m}H_{JC}\ket{\varphi_{i},n} = \varepsilon_{i,n}\delta_{i,j}\delta_{n,m}  \nonumber\\
        & + \lambda\mathcal{M}_{i,j}(\sqrt{n}\delta_{n,m+1}+\sqrt{n+1}\delta_{n+1,m}),
\end{align}
where $\varepsilon_{i,n} = E_{i}+\hbar\omega_{r}(n+1/2)$, $\mathcal{M}_{i,j} = \bra{\varphi_{i}}\frac{\partial H_0}{\partial \phi}\ket{\varphi_{j}}$ and $\lambda = (M/L)(2e/\hbar)\,\sqrt{\hbar\,/\,2\omega_r\,C}$. This interaction shifts the  frequency of the resonator and the intensity can be estimated by applying the perturbation theory on $\lambda$ up to the second order \cite{Park2017:PRB}:
\begin{equation}\label{EQN:FreqShift} 
    \delta w_{r} = -\sum_{i\neq j}\lambda^{2}|\mathcal{M}_{i,j}|^{2}\left(\frac{1}{ E_{i,j}/\hbar - \omega_r} - \frac{1}{ E_{i,j}/\hbar + \omega_r}\right),
\end{equation}
where $E_{i,j} = E_{j} - E_{i}$.

The excitation spectrum consists of the energy difference between the ground state, with energy $E_{G} = E_{i} + \hbar\omega_r/2$, for $E_{i}<0$, and an excited state, $\ket{\varphi_{i},0}$ for $E_{i}>0$ and $\ket{\varphi_{i},n}$ for $n\neq 0$. To first order, a photon is absorbed when the drive frequency, $f_d$, satisfies $\hbar f_d = \Delta E$ up to a broadening. In the main text, numerical fits to the experimental data obtained by discretizing and diagonalizing the Jaynes-Cummings Hamiltonian in \cref{EQN:JC} are discussed.

\section*{Appendix J: SOC Strength and JC interaction strength Dependence}

\begin{figure}[ht!]
\centering
\includegraphics[width=\columnwidth]{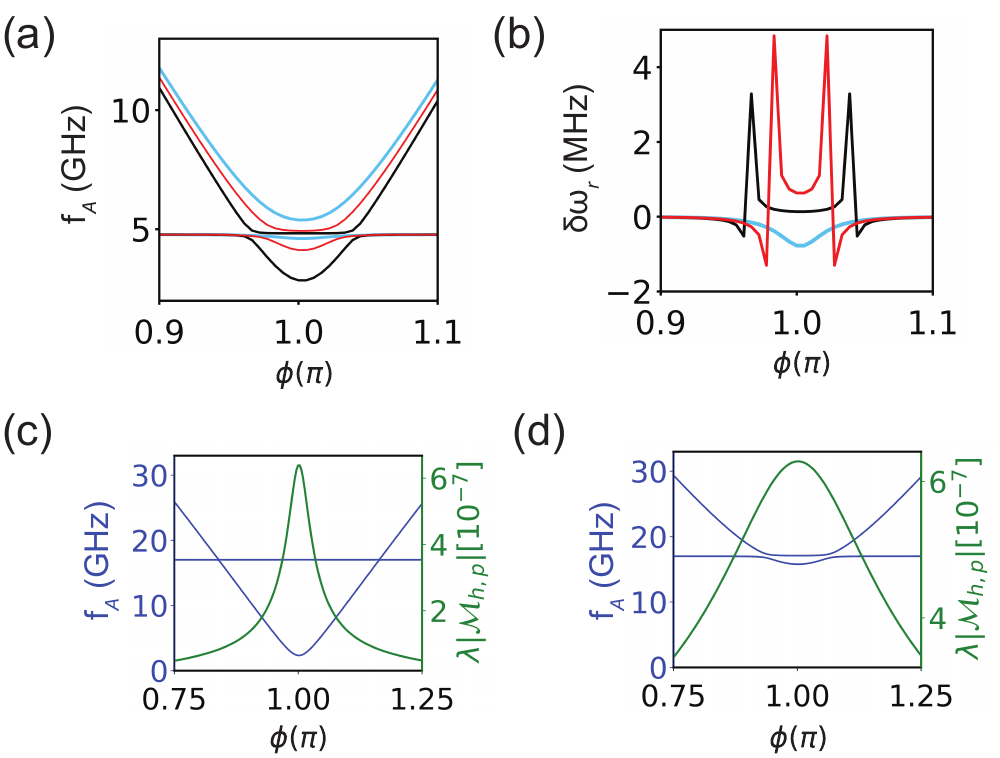}
\caption{(a) Excitation lines and (b) shift in the resonance frequency for $\alpha=\SI{2.95}{\milli eV}$ (red line), $\alpha=\SI{7.00}{\milli eV}$ (black line) and $\alpha=\SI{14.00}{\milli eV}$ (green line). (c) and (d) Numerical results for excitation lines (blue) and resonator-junction interaction strengths (green) for (c) $L_{con}=\SI{31}{\nano m}$ and for (d) $L_{con}=\SI{38}{\nano m}$. The parameters used in (a)-(b) are $m^{*}=0.036\,m_0$, $\Delta_0=\SI{0.15}{\milli eV}$, $\mu = \SI{15}{\milli eV}$ and $f_r=\SI{4.77}{\giga Hz}$, with the effective confinement length $L_{con}=\SI{32}{\nano m}$.}
\label{FIG:theory_2} 
\end{figure}

In \cref{FIG:theory_2}(a) and (b), we present the excitation lines and the shift of the resonance frequency $\delta w_{r}$ (see \cref{EQN:FreqShift}) for $\alpha$ at $\alpha=\SI{2.95}{\milli eV}$, $\alpha=\SI{7.00}{\milli eV}$ and $\alpha=\SI{14.00}{\milli eV}$. We see that for different Rashba SOC strengths, the excitation lines are shifted, resulting in different positions of anticrossings as seen in \cref{FIG:theory_2}(a). The shift in the position of the anticrossings are also seen in the spikes in $\delta w_{r}$, depicted in \cref{FIG:theory_2}(b). We note that tuning the Rashba SOC strength could modify the appearance (or lack thereof) of an anticrossing for a given chemical potential $\mu$ as shown in \cref{FIG:theory_2}(b). However, these anticrossing positions also have a strong dependence on $\mu$, as discussed in the main text. 

Further, we consider the effect of the Jaynes-Cummings interaction strength $\lambda\, |\mathcal{M}_{h,p}|$ between the particle and hole ABS states in \cref{FIG:theory_2}(c) and (d). The excitation lines (blue lines) are obtained using different confinement length, $L_{con}=\SI{31}{\nano m}$ (\cref{FIG:theory_2}(c)) and $L_{con}=\SI{38}{\nano m}$ (\cref{FIG:theory_2}(d)). We see that, for ABS levels relatively far from zero, the anticrossing between the resonator mode (horizontal line) and the junction modes is prominent as seen in \cref{FIG:theory_2}(d), which is accompanied by a high interaction strength (green line). In contrast, the anticrossing in \cref{FIG:theory_2}(c) is barely visible, accompanied by a low interaction strength.

\bibliography{refs}

\end{document}